\documentclass[prd,twocolumn,amsmath,amsfonts,amssymb,showpacs,nofootinbib]{revtex4}
\usepackage{epsfig,graphicx,bm,color}
\begin{document}
\title{Thermal decoupling and the smallest subhalo mass\\ in  dark matter models with Sommerfeld-enhanced annihilation rates}
\author{Laura~G.~van~den~Aarssen}
\email{laura.van.den.aarssen@desy.de}
\affiliation{{II.} Institute for Theoretical Physics, University of Hamburg, 
Luruper Chausse 149, DE-22761 Hamburg, Germany}
\author{Torsten~Bringmann}
\email{torsten.bringmann@desy.de}
\affiliation{{II.} Institute for Theoretical Physics, University of Hamburg, 
Luruper Chausse 149, DE-22761 Hamburg, Germany}
\author{Ya\c sar~C.~Goedecke}
\email{yasar.goedecke@desy.de}
\affiliation{{II.} Institute for Theoretical Physics, University of Hamburg, 
Luruper Chausse 149, DE-22761 Hamburg, Germany}

\date{June 14, 2012}


\begin{abstract}
We consider dark matter consisting of weakly interacting massive particles (WIMPs) and revisit in detail its thermal evolution in the early universe, with a particular focus on models where the annihilation rate is enhanced by the Sommerfeld effect.
After chemical decoupling, or freeze-out, dark matter no longer annihilates but is still kept in local thermal equilibrium due to scattering events with the much more abundant standard model particles. During kinetic decoupling, even these processes stop to be effective, which eventually sets the scale for a small-scale cutoff in the  
matter density fluctuations. Afterwards, the WIMP temperature decreases 
more quickly than the heat bath temperature, which causes dark matter to reenter an era of 
annihilation if the cross-section is enhanced by the Sommerfeld effect. Here, we give a detailed and self-consistent description of these effects. As an application, we consider the phenomenology of simple  leptophilic models that have been discussed in the literature and find that the relic abundance can be affected by as much two orders of magnitude or more. We also compute the mass of the smallest dark matter subhalos in these models and find it to be in the range $\mathcal{O}(10^{-10}M_\odot)-\mathcal{O}(10\,M_\odot)$; even much larger cutoff values are possible if the WIMPs couple to force carriers lighter than about 100\,MeV. 
We point out that a precise determination of the cutoff mass
allows to infer new limits on the model parameters, 
in particular from gamma-ray observations of galaxy clusters, that are highly complementary to existing constraints from $g-2$ or beam dump experiments.

\end{abstract}

\maketitle

\newcommand{\be}{\begin{equation}}
\newcommand{\ee}{\end{equation}}
\newcommand{\bea}{\begin{eqnarray}}
\newcommand{\eea}{\end{eqnarray}}
\newcommand{\eref}[1]{Eq.~\eqref{#1}}

\hyphenation{}

\section{Introduction}

A plethora of independent cosmological observations, taken over a large range of distance scales, strongly supports the existence of cold, non-baryonic dark matter (DM); according to the most recent estimates, it contributes a fraction of $\Omega_{\rm DM} = 0.229 \pm 0.015$ to the total energy content of the universe \cite{Komatsu:2010fb}. Weakly interacting massive particles (WIMPs) are the leading candidate for the so far still obscure nature of the DM, well motivated from particle physics and potentially offering promising prospects for detection in indirect, direct and accelerator searches \cite{DM_reviews}. When mentioning DM in this article, we will always refer to WIMPs.

One of the most appealing aspects of WIMPs is their thermal production in the early universe, resulting in a relic density that is roughly of the same order as the observed DM density today \cite{WIMP_old}. While this  somewhat simplistic picture actually changes when taking into account complications \cite{Griest:1990kh} like co-annihilations, thresholds or resonances,
the main idea still remains the same: at very high temperatures, WIMPs are kept in thermal equilibrium through annihilation and creation processes with the standard model particles of the heat bath; once the interaction rate  falls considerably behind the Hubble expansion, the WIMP number density ``freezes out'' and is only affected by the further expansion of the universe  (for a very thorough treatment, see e.g.~Refs.~\cite{Gondolo:1990dk,Edsjo:1997bg}).

Even after this \emph{chemical decoupling}, WIMPs stay in thermal contact with the heat bath due to the much more frequent elastic-scattering events with the relativistic standard model particles  \cite{Schmid:1998mx,Chen:2001jz}. At \emph{kinetic decoupling}, which actually happens on a rather short timescale \cite{Bringmann:2006mu,Bringmann:2009vf}, even these scattering processes cease to be effective; 
in the standard picture, there is then no further interaction between WIMPs and the heat bath. From this point on,  density perturbations in the DM component are thus no longer tightly bound to density perturbations in the radiation component and can freely develop, with free-streaming effects setting a lower bound to the size of over-dense regions that will eventually collapse under the influence of gravitation \cite{Hofmann:2001bi,Berezinsky:2003vn,Green:2003un,Green:2005fa};  acoustic oscillations lead to a similar cutoff \cite{Loeb:2005pm,Bertschinger:2006nq}. 

While the processes that determine the relic density and the cutoff in the power spectrum of matter density fluctuations  often happen at completely unrelated timescales, however, this does not necessarily have to be the case. In models where the annihilation rate at small relative velocities is strongly enhanced, e.g., the effects of annihilation and  scattering processes on the DM distribution can strongly interfere and change the simple picture sketched above. In particular, one may encounter a new era of DM annihilations \emph{after} kinetic decoupling has taken place \cite{Dent:2009bv,Zavala:2009mi,Feng:2010zp}. Physically, such strongly enhanced annihilation rates can be motivated by the repeated exchange of relatively light force carriers, known as the Sommerfeld effect \cite{Sommerfeld:1931}; in the context of DM annihilation, this has first been studied for heavy neutralino DM coupling to $SU(2)$ gauge bosons \cite{Hisano:2002fk,Hisano:2003ec,Hisano:2004ds}, but seen a boost of interest (see, e.g.~Refs.~\cite{Cirelli:2008jk,ArkaniHamed:2008qn,Pospelov:2008jd,Cholis:2008qq,Nomura:2008ru,Fox:2008kb,Lattanzi:2008qa}) in the context of possible DM explanations for the apparent anomalies in the observed cosmic ray electron and positron fluxes \cite{CR_excess}.

Here, we revisit in detail how WIMPs decouple from the thermal bath in the early universe and derive, by extending the standard calculation, a coupled system of equations that  describes the evolution of the WIMP number density and velocity dispersion (aka the WIMP ``temperature"). This allows  to compute the evolution of these quantities to a very high precision and, in particular, treat situations like the one mentioned in the previous paragraph in a fully consistent way that can be applied to any DM model where the annihilation rate is enhanced for small relative velocities. 

As an application, we  focus on a simple toy-model with large Sommerfeld enhancement and compute both the final relic density and the size of the smallest subhalo masses. We demonstrate that the above-mentioned effects can be sizable and that the consistent treatment derived here is indeed necessary to reliably compute these quantities in  DM models where the Sommerfeld effect is relevant. Interestingly, gamma rays from DM annihilation in galaxy clusters may place strong lower limits on the cutoff mass in these kind of models \cite{Pinzke:2009cp,Pinzke:2011ek}; we demonstrate here that a reliable and self-consistent computation of this quantity can therefore be used, in principle, to translate these limits into constraints on the parameter space that are highly complementary to existing constraints from $g-2$ or beam dump experiments.

This article is organized as follows. We start in Sec.~\ref{sec:decoupling} with a concise review of chemical and kinetic decoupling of WIMPs, extending the standard treatment in such a way as to take into account possible interferences between the two. In Sec.~\ref{sec:new_ann}, we then discuss in detail the evolution of the DM phase-space distribution after kinetic decoupling, with a particular focus on the possibility of Sommerfeld-enhanced annihilation rates. Section \ref{sec:models} is devoted to the application of our general formalism to a concrete leptophilic DM particle model.
We discuss our results in Sec.~\ref{sec:disc} and present our conclusions and an outlook in Sec.~\ref{sec:conc}. In Appendix \ref{app:sommerfeld}, we collect some relevant details about the Sommerfeld effect. For the leptophilic toy-model studied in Sec.~\ref{sec:models}, we provide current constraints in Appendix \ref{app:Constraints} and, for convenience, a collection of scattering and annihilation matrix elements in Appendix \ref{app:m2}.

\section{The standard thermal evolution of WIMPs}
\label{sec:decoupling}

In this section, we review the standard case of thermally produced particles in the  expanding universe. The evolution of their phase-space density $f(\mathbf{p})$, in particular, is described by the Boltzmann equation which in a Friedmann-Robertson-Walker metric reads (see, e.g., \cite{Bernstein:1988bw,Kolb})
\be
  \label{boltzmann}
  E\left(\partial_t-H\mathbf{p}\cdot\nabla_\mathbf{p}\right)f=C[f]\,.
\ee
Here, $p^\mu=(E,\mathbf{p})$ denote comoving WIMP momenta and $H=\dot a/a$ the Hubble parameter; all interactions are contained in the collision term on the right-hand side. As we will see in detail, this equation describes how WIMPs start in perfect thermodynamic equilibrium with the early, very dense and hot universe and eventually decouple completely from the heat bath, in several distinct stages, as the universe continues to expand.

\subsection{Chemical decoupling}
\label{sec:cd}

The evolution of the DM particle number density
\be
  n_\chi\equiv g_\chi \int\frac{d^3p}{(2\pi)^3}f(\mathbf{p})
\ee  
   is affected both by the expansion of the universe and by DM annihilation into, or creation from, SM particles. Restricting ourselves to 2-body processes, the corresponding contribution to the collision term is given by
\bea
  \label{Cfull}
  C_{\rm ann}&=&\frac{1}{2g_\chi}\sum_X\int\frac{d^3k}{(2\pi)^32\omega}\int\frac{d^3\tilde k}{(2\pi)^32\tilde \omega}\int\frac{d^3\tilde p}{(2\pi)^32\tilde E}\nonumber\\
  &&\times (2\pi)^4\delta^{(4)}(\tilde p+p-\tilde k-k)\nonumber\\
&&\left[
\left|\mathcal{M}\right|^2_{\bar\chi\chi\leftarrow \bar XX}g(\omega)g(\tilde \omega)
-\left|\mathcal{M}\right|^2_{\bar\chi\chi\rightarrow \bar XX}f(E)f(\tilde E)
\right]\nonumber\\
&=&  g_\chi E\sum_X\int\frac{d^3\tilde p}{(2\pi)^3}\,v_{\rm rel} \sigma_{\bar\chi\chi\rightarrow \bar XX}\nonumber\\
&&\times\left[
f_{\rm eq}(E)f_{\rm eq}(\tilde E)-f(E)f(\tilde E)
\right]\,.
\eea
where  $k^\mu=(\omega,\mathbf{k})$ 
and $\tilde k^\mu=(\tilde \omega,\tilde{\mathbf{k}})$ are the 4-momenta of the SM particles $X$ and $g\!=\!g_{\rm eq}\!=\!\left(e^{\omega/T}\pm1\right)^{-1}$ their distribution functions; 
$\left|\mathcal{M}\right|^2$ refers to the matrix element summed over both SM  and DM internal (spin) degrees of freedom $g_{X,\chi}$.
Note that we can safely neglect Pauli-suppression factors in the nonrelativistic regime ($\langle\mathbf{p}^2\rangle\ll m_\chi^2$) we are considering here. The second step follows from $CP$ invariance and the fact that annihilation and creation processes should happen with the same frequency in  equilibrium; the velocity appearing here is the M\o ller velocity, $v_{\rm rel}=v_{\rm M\o l}\equiv ({E \tilde E})^{-1}{\sqrt{(p \cdot \tilde p)^2-m_\chi^4}}$.

In order to proceed, one usually assumes that $f(E)\propto f_{\rm eq}(E)= e^{-E/T}$, with a factor of proportionality that describes an effective chemical potential which may depend on $T$ (but not $E$); this is motivated by the fact that the much more abundant scattering processes of DM with SM particles still keep the DM particles in kinetic, but not chemical equilibrium -- see also the following Section.
Integrating \eref{boltzmann} over $\int{d^3p}\, g_\chi/\left[{(2\pi)^3 E}\right]$ then results in\footnote{
When including coannihilations \cite{Griest:1990kh}, this equation takes the same form -- with \eref{thermalsv} being replaced by an \emph{effective} thermally averaged cross-section $\langle \sigma_{\rm eff} v\rangle$ 
and $n_\chi^{\rm (eq)}$ denoting the total (equilibrium) number density of \emph{all} co-annihilating particles
\cite{Edsjo:1997bg}.
} 
\be
  \label{boltz_firstmoment_g}
 \dot n_\chi+3Hn_\chi=-\langle\sigma v_{\rm rel}\rangle_{\rm eq}\left(n_\chi^2-{n_\chi^{\rm eq}}^2\right)\,,
\ee
with $n_\chi^{\rm eq}=m_\chi^3g_\chi K_2(x)/(2\pi^2x)$ in the nonrelativistic regime and
\bea
  \label{thermalsv}
  \langle \sigma v_{\rm rel}\rangle_{\rm eq}&\equiv&\frac{g_\chi^2}{{n_\chi^{\rm eq}}^2}\int\frac{d^3p}{(2\pi)^3}\int\frac{d^3\tilde p}{(2\pi)^3}\nonumber\\
  &&\times v_{\rm rel} \sigma_{\bar\chi\chi\rightarrow \bar XX}f_{\rm eq}(E)f_{\rm eq}(\tilde E)\\
&\simeq& \frac{4x^{3/2}}{\sqrt{\pi}}\int_0^1\left(\sigma v_{\rm rel}\right)v^2e^{-v^2x}dv\,,\label{svavsim}
\eea
where  $K_2$ is the modified Bessel function of second order and $v$ is the  velocity of \emph{each} WIMP in the center-of-mass frame. The last approximation is valid for large $x$ and, for $x\gtrsim10$,  reproduces the full analytical result \cite{Gondolo:1990dk}  to an accuracy of better than 1\% for all relevant functional dependences of $\left(\sigma v_{\rm rel}\right)$. A simple power-law
\be
  \label{svexpand}
  \left(\sigma v_{\rm rel}\right)=\sigma_0v^{2n}\,,
\ee
in particular, results in
\be
  \langle \sigma v_{\rm rel}\rangle_{\rm eq}\simeq \frac{2\sigma_0}{\sqrt{\pi}}\,\Gamma\!\left(n+\frac32\right)x^{-n}\equiv\tilde\sigma_0x^{-n}\,.
\ee

The next step is to introduce  dimensionless variables 
\bea
  x&\equiv& m_\chi/T\,,\\
  Y&\equiv& n_\chi/s\,, \label{Ydef}
\eea
 where $T$ is the temperature of the heat bath and $s = g_{*\text{S}}(T) \frac{2 \pi^{2}}{45} T^{3}$
the entropy density. Using entropy conservation, $\partial_t\left(a^3s\right)=0$, one can then  transform \eref{boltz_firstmoment_g} to the convenient form
\be
    \label{dYdx}
	\frac{Y'}{Y} = - \left(1-\frac{x}{3} \frac{g_{*\text{S}}'}{ g_{*\text{S}}}\right)
 \frac{n_\chi  \langle\sigma v_{\rm rel} \rangle_{\rm eq} }{Hx} \left(1-\frac{Y^{2}_{\text{eq}}}{Y^2}\right)\,, 
 \ee
where $'\!\!\equiv d/dx$. Note that without the assumption of $f(E)\propto f_{\rm eq}(E)$, we would have arrived at a very similar equation, with $\big(1-Y_{\rm eq}^2/Y^2\big)$ being replaced by $\big(\langle \sigma v_{\rm rel}\rangle/\langle \sigma v_{\rm rel}\rangle_{\rm eq}-Y_{\rm eq}^2/Y^2\big)$, where $\langle \sigma v_{\rm rel}\rangle$ is defined exactly as in \eref{thermalsv}, but for an arbitrary WIMP distribution, i.e.~with $(f_{\rm eq},n_\chi^{\rm eq})\rightarrow(f,n_\chi)$.
During radiation domination, $H^2\!=\!\frac{4\pi^3}{45m_{\rm Pl}^2}g_{\rm eff}T^4$, Eq. (\ref{dYdx}) simplifies further to 
\begin{equation}
         \label{dYdxsimp}
	\frac{dY}{dx} = - \lambda x^{-2-n}  (Y^{2}-Y^{2}_{\text{eq}})\,,
\end{equation}
where we have assumed \eref{svexpand} to hold and introduced
\be
 \label{lambdadef}
  \lambda\equiv \frac{g_{*\text{S}}}{\sqrt{g_{\text{eff}}}}  \left(1-\frac{x}{3} \frac{g_{*\text{S}}'}{ g_{*\text{S}}}\right)
  \sqrt{\frac{\pi}{45}}m_{\rm Pl}m_\chi \tilde\sigma_0\,.
\ee

At very early times,  \eref{dYdx} forces $Y$ to follow the equilibrium value $Y=Y_{\rm eq}$, i.e.~WIMPs are  very efficiently kept in chemical equilibrium with the heat bath and detailed balance is maintained between DM annihilation and  creation from  heat bath particles. At \emph{chemical decoupling}, the factor in front of $(1-Y_{\rm eq}^2/Y^2)$ in \eref{dYdx} has dropped to a value where DM annihilations are no longer efficient enough to maintain chemical equilibrium and  $Y$ starts to decrease at a slower rate than $Y_{\rm eq}$. From inspection of this equation, this roughly happens when $H\sim\langle\sigma v_{\rm rel}\rangle_{\rm eq}\, n_\chi^{\rm eq}$; for WIMPs  with the right relic density to explain all of the observed DM  today, this is the case at $x_{\rm cd}\simeq20-28$ (see, e.g., \cite{Bringmann:2009vf}). Only slightly later, the DM density freezes out completely, i.e.~$Y$ (which roughly corresponds to its comoving number density)   stays constant -- for standard WIMPs essentially until today. 
In the case of a Sommerfeld-enhanced annihilation rate, the expected range of $x_{\rm cd}$ essentially does not change. Compared to the standard case, however, the final stage of the freeze-out process may be delayed. For the models that we will consider here, this corresponds to an increase in $x_{99}$ by up to a factor of $20$, or even more in the case of resonances, where we define $x_{99}$ as the value of $x$ when $Y$ differs from its asymptotic value by less than 1\%.

\subsection{Kinetic decoupling}

After chemical decoupling, WIMPs are still kept in local thermal equilibrium with the heat bath by the much more frequent elastic scattering processes with SM particles.\footnote{
See Ref.~\cite{Arcadi:2011ev} for an example of how to treat even inelastic-scattering processes
in case there exists another heavy particle highly degenerate in mass with the DM particle $\chi$.} These contributions to the collision term read:
\bea
  C_{\rm el} &=& \frac{1}{2 g_\chi}\sum_X\int\frac{d^3k}{(2\pi)^32\omega}\int\frac{d^3\tilde k}{(2\pi)^32\tilde \omega}\int\frac{d^3\tilde p}{(2\pi)^32\tilde E}\nonumber\\
  &&\times(2\pi)^4\delta^{(4)}(\tilde p+\tilde k-p-k){\left|\mathcal{M}\right|}^2_{\chi X\leftrightarrow \chi X}
  \nonumber\\
  &&
  \Big[\left(1\mp g^\pm(\omega)\right)\, g^\pm(\tilde\omega)f(\mathbf{\tilde p})  \nonumber\\
  &&\quad-\left(1\mp g^\pm(\tilde\omega)\right)\, g^\pm(\omega)f(\mathbf{p})\Big]\,,
\eea
where 4-momenta with (without) a tilde describe ingoing (outgoing) particles and ${\left|\mathcal{M}\right|}^2$ again is the matrix element squared and summed over all spin states. Note that  $C_{\rm el}$ only contains particle number-conserving  processes,  so it  does not contribute to \eref{boltz_firstmoment_g} -- as can straight-forwardly be checked explicitly \cite{Bringmann:2006mu}.

Rather than the first moment  of the Boltzmann equation, as in \eref{boltz_firstmoment_g} for the determination of the chemical freeze-out temperature, one may  consider its second moment  to get an  accurate description of when the DM particles leave thermal equilibrium with the heat bath \cite{Bringmann:2006mu,Bringmann:2009vf}. 
To this end, it is very convenient to introduce the parameter
\be
  T_\chi\equiv\frac{g_\chi}{3\,m_\chi n_\chi}\int\frac{d^3p}{(2\pi)^3}\mathbf{p}^2f(\mathbf{p})\,,
\ee
which would correspond to the temperature of a nonrelativistic WIMP \emph{if} $f$ were a thermal distribution. The difference between $T_\chi$ and $T$ thus indicates  how well the WIMPs are kept in thermal equilibrium with the heat bath.
In analogy to \eref{Ydef}, we further introduce the dimensionless quantity
\be
\label{ydef}
  y\equiv \frac{m_\chi T_\chi}{s^{2/3}}\,.
\ee
Multiplying \eref{boltzmann} by $g_\chi\mathbf{p}^2/E$, integrating it over $\mathbf{p}$ and keeping only the leading order terms in $\mathbf{p}^2/m_\chi^2$ then leads, after a somewhat lengthy calculation, to
\begin{widetext}
\be
  \label{dydx}
  \frac{y'}{y}=- \frac{Y'}{Y}\left(1-\frac{\langle \sigma v_{\rm rel}\rangle_2}{\langle \sigma v_{\rm rel}\rangle}\right)
               - \left(1-\frac{x}{3} \frac{g_{*\text{S}}'}{ g_{*\text{S}}}\right)
  \frac{2m_\chi c(T)}{Hx}\left(1-\frac{y_{\rm eq}}{y}\right)\,,
\ee
with
\bea
\langle \sigma v_{\rm rel}\rangle_2&\equiv&\frac{g_\chi^2}{3 T m_\chi n_\chi^2}\int\frac{d^3p}{(2\pi)^3}\int\frac{d^3\tilde p}{(2\pi)^3}
p^2 \left(v_{\rm rel} \sigma_{\bar\chi\chi\rightarrow \bar XX}\right)f(E)f(\tilde E)\\
  &\stackrel{f\propto e^{-E/T}}{=}&
 \frac{x}{3K_2^2(x)}\int_{1}^\infty\!\!\left(\sigma v_{\rm rel}\right)\,\sqrt{\tilde s-1}(2\tilde s-1)
 \left[ 2(\tilde s-1)K_1(2x\sqrt{\tilde s})
 +\tilde s^{-\frac12}(4\tilde s-1)K_2(2x\sqrt{\tilde s})
  \right]\,d\tilde s \label{svav2eq}\\
&\simeq& \frac{2x^{3/2}}{\sqrt{\pi}}\int_0^1\left(\sigma v_{\rm rel}\right)v^2\left(1+\frac{2}{3}xv^2\right)e^{-v^2x}dv\,,\label{svav2}
\eea
and \cite{Bringmann:2009vf}
\be
  \label{cTdef}
  c(T) =  \frac{1}{12(2\pi)^3m_\chi^4T} \sum_X
   \int dk\,k^5 \omega^{-1}\,g^\pm\left(1\mp g^\pm\right)\mathop{\hspace{-12ex}{\left|\mathcal{M}\right|}^2_{t=0}}_{\hspace{4.5ex}s=m_\chi^2+2m_\chi\omega+m_X^2}\,.
\ee
\end{widetext}
In arriving at \eref{dydx}\footnote{
This expression improves  the corresponding Eq.~(10) in Ref.~\cite{Bringmann:2009vf} by using a more suitable definition of $y$; more importantly, we allow here explicitly for the case that DM annihilation has not ended yet (i.e.~$dY/dx\neq0$). 
}, we have assumed that $Y_{\rm eq}\ll Y$ and \eref{svav2eq} is valid if $f(E)\propto e^{-E/T}$ ($\tilde s$ is the dimensionless version of the Mandelstam variable $s\equiv4m_\chi^2\tilde s$).
The approximation given in \eref{svav2} takes a form very similar to \eref{svavsim} and also exhibits an accuracy that is very similar, i.e.~much better than at the percent-level for the values of $x$ that we are interested in here.   For a scaling like in \eref{svexpand}, $\sigma v_{\rm rel}\propto v^{2n}$, in particular, we find
\be
\frac{\langle \sigma v_{\rm rel}\rangle_2}{\langle \sigma v_{\rm rel}\rangle}=1+\frac{n}{3}\,.\label{annrate_ratio}
\ee

Note that, for $Y'=0$, \eref{dydx} is the exact analogue of \eref{dYdx}: as long as the scattering processes are frequent enough, $y$ follows the heat-bath value $y_{\rm eq}\equiv m_\chi Ts^{-2/3}$, i.e.~we have $T_\chi=T$ as expected. At very late times, on the other hand, the factor in front of $\big(1-y_{\rm eq}/y\big)$ becomes vanishingly small and $y$ stays constant, i.e.~$T_\chi\propto a^{-2}$, 
 which simply reflects the redshift of the WIMP momenta  due to the expansion of the universe.
The fact that the transition between these two regimes happens on a rather short timescale \cite{Bringmann:2006mu,Bringmann:2009vf} allows to conveniently define the temperature of \emph{kinetic decoupling} as
\be
  x_{\rm kd}=\frac{m_\chi}{T_{\rm kd}}\equiv \left.y\right|_{x\rightarrow\infty}^{Y'\stackrel{!}{=}0}
  \times\left.\frac{s^{2/3}}{T^2}\right|_{T=T_{\rm kd}} \,.
\ee
As expected, kinetic decoupling happens considerably later than chemical decoupling; in the case of neutralino DM, e.g., one finds $x_{\rm kd}/x_{\rm cd}\sim10-4000$ (or $T_{\rm kd}\sim5\,{\rm MeV}-5\,{\rm GeV}$) \cite{Bringmann:2009vf}.

As we will see, things may change considerably in situations where we cannot actually neglect $Y'$. We therefore advocate, as indicated, to use the above definition of $x_{\rm kd}$ \emph{only after setting $Y'/Y\equiv0$ by hand in \eref{dydx}}. This definition then accurately reflects the intuitive meaning of kinetic decoupling even in the case where we cannot neglect $Y'$, i.e.~the point where scattering processes with heat bath particles are no longer effective.

\section{Evolution of dark matter density after kinetic decoupling}
\label{sec:new_ann}

In the conventional WIMP scenario, the  collision term in \eref{boltzmann} can be completely neglected by the time of kinetic decoupling, i.e.~the further evolution of $f$ is only governed by the  expansion of the universe -- at least until the tiny primordial  density fluctuations have grown large enough to trigger structure formation and self-annihilation may start again. For the case of Sommerfeld-enhanced annihilation rates, as we will discuss now in some detail, this part of the evolution history is qualitatively different and much more complex.

\subsection{A new era of annihilation}
\label{reenter_annihilation}
Let us focus on the standard situation where $x_{\rm kd}\gg x_{\rm cd}$; around and after kinetic decoupling, we thus have $Y\gg Y_{\rm eq}$. Therefore, the formal solution to \eref{dYdx} is given by:
\be
  \label{deltaY}
  Y(x)^{-1}=Y(x_i)^{-1}+\int_{x_i}^x
   \left(1-\frac{x}{3} \frac{g_{*\text{S}}'}{ g_{*\text{S}}}\right)
	  \frac{s  \langle\sigma v_{\rm rel} \rangle }{Hx} dx\,,
\ee
for any $x_i\gg x_{\rm cd}$. In order to gain some qualitative understanding of this expression, let us again assume that $\sigma v_{\rm rel}\propto v^{2n}$. As discussed in the previous Section, we 
roughly have $v\simeq p/m_\chi\propto x^{-1/2}$ before kinetic decoupling and  $v\propto x^{-1}$ afterwards; as a consequence, we expect  $\langle\sigma v_{\rm rel} \rangle\propto x^{-\tilde n}$, where 
\be
\tilde n = \left\{
 \begin{array}{ll}
 n & {\rm for}\, x\lesssim x_{\rm kd}\\
 2n & {\rm for}\, x\gtrsim x_{\rm kd}\\
 \end{array}
 \right.\,.
\ee
Approximating $\lambda$ given in \eref{lambdadef} to be constant, we can now integrate \eref{deltaY}
and find
\be
 Y(x)^{-1}-Y(x_i)^{-1}\simeq\lambda\left\{
 \begin{array}{ll}
 \frac{1}{1+\tilde n}\left(\frac{1}{x_i^{1+\tilde n}}-\frac{1}{x^{1+\tilde n}}\right) & {\rm for}\, \tilde n\neq-1\\
 \ln (x/x_i) &  {\rm for}\,\tilde n=-1 
 \end{array}
 \right.\,.
\ee
Clearly, an appreciable change in $Y$ for $x>x_i$ is only possible for $\tilde n\leq-1$; in fact, taken at face value, annihilations would never cease in that case. For the standard WIMP scenario, this is impossible to achieve since $s$-wave annihilation implies $n=\tilde n\!=\!0$ and  higher partial waves are even more strongly suppressed (e.g.~$n\!=\!1$ for the $p$-wave). 
For a Sommerfeld-like $1/v$ enhancement of $s$-wave annihilations, however, the situation looks very different and WIMPs may reenter an era of annihilation \cite{Dent:2009bv}: in this case, we  do have $\tilde n=-1$ after kinetic decoupling. On resonances, we could actually have $\langle\sigma v_{\rm rel}\rangle\propto v^{-2}$, i.e.~$\tilde n=-2$ (see Appendix \ref{app:sommerfeld}); note that this would imply a non-negligible annihilation rate even \emph{before} kinetic decoupling (with $n=\tilde n=-1$).

Let us now have a more detailed and quantitative look at this effect. Assuming that the DM velocity distribution stays Maxwellian even after kinetic decoupling (see the following Sec.~\ref{dm_selfscatter}), we can use Eqs.~(\ref{svavsim},\ref{svav2}) to calculate $\langle\sigma v_{\rm rel} \rangle_{(2)}$ simply by replacing $T\rightarrow T_\chi$.
For a Sommerfeld-enhanced $s$-wave annihilation, e.g., we then have
\be
  \langle\sigma v_{\rm rel} \rangle
  =  \left.\langle S(v)\sigma_0\rangle\right|_{T=T_\chi} 
  \simeq  2\sqrt{\frac{m_\chi}{\pi T_\chi}}\sigma_0\,,
\ee
where  the last step is valid if velocities of the order of $v\sim\bar{v}\equiv\sqrt{8T_\chi/\pi m_\chi}$  fall into the Coulomb regime where $S(v)\propto v^{-1}$;
this is  exactly the $T_\chi^{-1/2}\propto x^{1/2}$ scaling mentioned above.
For a full understanding of the evolution of the WIMP number density and temperature in this regime, however, we need to solve the following coupled system of differential equations for $y$ and $Y$ that follows from Eqs.~(\ref{dYdx}, \ref{dydx}):
\bea
	\frac{Y'}{Y} &=& - \frac{1-\frac{x}{3} \frac{g_{*\text{S}}'}{ g_{*\text{S}}}}{Hx}
	 sY \left.\langle\sigma v_{\rm rel} \rangle\right|_{x=m_\chi^{2}/(s^{2/3}y)}  \label{Y_coupled}\\
  \frac{y'}{y}&=&-\frac{1-\frac{x}{3} \frac{g_{*\text{S}}'}{ g_{*\text{S}}}}{Hx}
	  \Bigg[2m_\chi c(T)\left(1-\frac{y_{\rm eq}}{y}\right)  \label{y_coupled}\\ &&  
	  \qquad\qquad- sY  \Big(\langle\sigma v_{\rm rel} \rangle-\langle\sigma v_{\rm rel} \rangle_2\Big)_{x=m_\chi^{2}/(s^{2/3}y)}
	 \Bigg]\,. \nonumber 
\eea
This set of equations provides one of our central results; it clearly demonstrates that kinetic and chemical decoupling cannot, in general, be treated separately. 

Some insight into the asymptotic behavior of these coupled equations is achieved by considering 
 the limit  where $x\gg x_{\rm kd}$, i.e.~where the scattering term proportional to $c(T)$ can be neglected. 
Assuming again $\sigma v_{\rm rel}\propto v^{2n}$, and using \eref{annrate_ratio}, we then find
\be
	 \frac{y'}{y} \simeq \frac{n}{3} \frac{Y'}{Y}=  \frac{\tilde n}{6} \frac{Y'}{Y}\,. \label{eq:asymprelationyY}
\ee
For $n<0$,  a decreasing  $Y$ will thus have the effect of increasing $y$ even after kinetic decoupling; this simply reflects the fact that the DM phase-space density is depleted of low velocity particles, thereby increasing the average velocity.

In the remainder of this Section, we will continue our discussion of the further evolution of $Y$ and $y$ on a rather general level; in Sec.~\ref{sec:models}, we will then consider a concrete class of WIMP DM models and
show that the effects discussed here can, indeed, be quantitatively quite important in determining the relic density or the small-scale cut off in the mass-distribution of DM subhalos.

\subsection{Dark matter self-scattering}
\label{dm_selfscatter}

In the presence of a Sommerfeld-enhanced annihilation rate, also the DM self-scattering rate is enhanced (see Appendix \ref{app:somm_scatt}); as a result, WIMPs can have a Maxwellian velocity distribution, 
\be
 f_v(v) = \sqrt{\frac{2}{\pi}} \left( \frac{m_{\chi}}{T_{\chi}} \right)^{3/2} v^{2} e^{-\frac12 m_{\chi} v^{2}/T_{\chi}}\, , 
 \ee
 even after kinetic decoupling has taken place \cite{Feng:2009hw, Buckley:2009in, Feng:2010zp}. In this case, as already indicated, we can easily evaluate the thermal averages that appear in Eqs.~(\ref{Y_coupled}, \ref{y_coupled}) by using the expressions in Eqs.~(\ref{svavsim}) and Eq.~(\ref{svav2}) with $T\rightarrow T_\chi$. We thus would like to estimate when the DM self-scattering ceases to be effective and  the DM velocity distribution starts to deviate from a Maxwellian form.

The self-scattering rate by which the velocities change by $\mathcal{O}(1)$ is given by \cite{Feng:2010zp, Feng:2009hw}
\begin{align}
	\Gamma_{s} &= n_{\chi} \langle  \sigma_{T} v_{\text{rel}} \frac{v^{2}_{\text{rel}}}{v^{2}_{0}} \rangle \,,\nonumber \\
	&=  \frac{g^{2}_{\chi}}{n_{\chi}} \int \frac{d ^{3}p}{2 \pi^{3}}   \frac{d ^{3}\tilde{p}}{2 \pi^{3}}  \ \sigma_{T} v_{\text{rel}} \frac{v^{2}_{\text{rel}}}{v^{2}_{0}} \ f(E) f(\tilde{E}) \,, \nonumber \\
	&\simeq \frac{16 n_{\chi}}{\sqrt{\pi}} \left(\frac{m_{\chi}}{T_{\chi}}\right)^{5/2} \int_{0}^{1} \sigma_{T} \frac{v^{5}}{\left(1+v^{2}\right)^{3}} e^{-v^{2}\tfrac{ m_{\chi}}{T_{\chi}}}  dv\,,  \label{eq:scatrate}
\end{align}
where in the last step $f\propto  e^{-E/T_{\chi}} $ was used. Here, $v_{0} = \sqrt{2 T_{\chi}/m_{\chi}}$ is the most probable velocity and the transfer cross section $\sigma_T$ is introduced in \eref{eq:transferxsection}. 

\begin{figure}[t]
\flushright
	\includegraphics[width=\columnwidth]{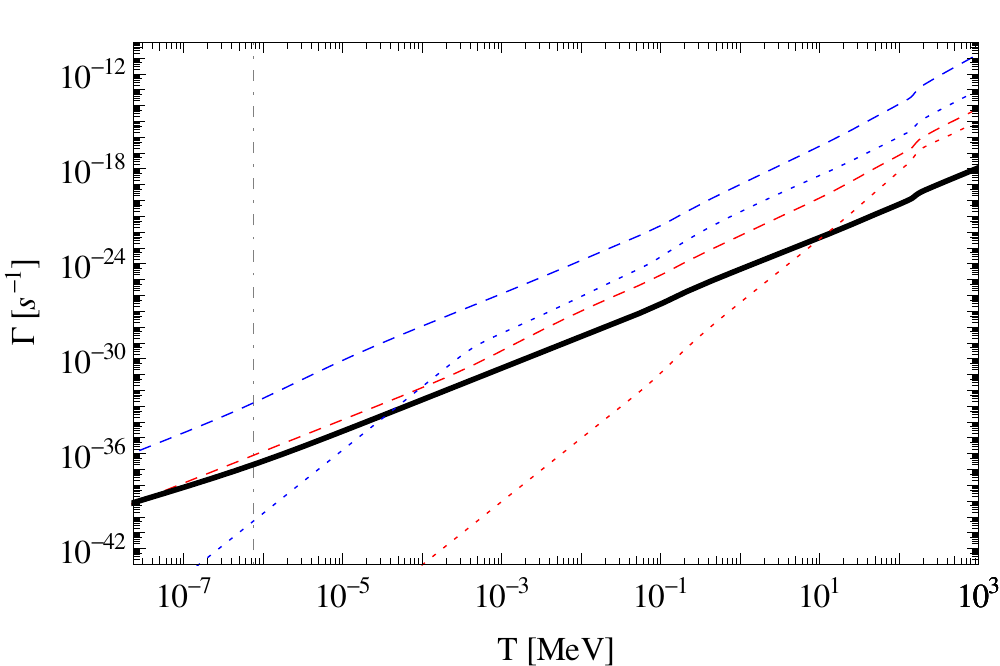}
	 \caption{The Hubble rate (full, black) in comparison with the effective DM self-scattering rate, introduced in \eref{eq:scatrate}, for a few extreme parameter sets (minimal in red, maximal in blue) of the model considered in Sec.~\ref{sec:models} (see also Table \ref{tab:selfscatter}). The dashed and dotted lines show the on- and off-resonant case, respectively. The dash-dotted line indicates, for comparison, the temperature of matter-radiation equality. The temperature $T_{\rm nt}$ below which the DM velocity distribution starts to  deviate from a Maxwellian form is roughly given by $\Gamma_{\text{s}}(T_{\text{nt}}) \sim H (T_{\text{nt}})$. } \label{fig:SSrate}
\end{figure}

Following Ref.~\cite{Feng:2010zp}, we define the temperature $T_{\text{nt}}$ at which the DM velocity distribution becomes nonthermal as the temperature where the scattering rate $\Gamma_{\text{s}}$ becomes comparable to the Hubble expansion rate 
\begin{equation}
	\Gamma_{\text{s}}(T_{\text{nt}}) \equiv H (T_{\text{nt}}).
\end{equation}
We note that a more precise determination of $T_{\text{nt}}$ would in principle be possible by  solving the second moment of the Boltzmann equation with a collision term that describes  WIMP self-interactions instead of scattering on SM particles. Unfortunately, however, one can no longer  use the fact that the momentum transfer is small in this case and a substantial extension to the formalism presented in Refs.~\cite{Bringmann:2006mu, Bringmann:2009vf} would be required, which is beyond the scope of this article.

 \begin{table}[t!]
 \centering
 \begin{tabular}{l||c|c|c|c||c|c}
      & $m_\chi$ & $m_\phi$ & $\alpha$ & $g_\ell$ &  $T_{\rm kd}$ & $T_{\rm nt}$\\
      & [TeV] &[GeV]&  &  & [MeV] & [MeV]\\
 \hline
  \hline
  Off-res. (min) & 5  & 5 & 0.0955 & $10^{-7}$ &  400 & $11.5$ \\
 \hline
  Off-res. (max) & 0.5 & 0.1 & 0.0154 & $10^{-1}$ & $0.0665$ &$4\times10^{-5}$ \\
  \hline
  On-res. (min) & 5 & 5 & 0.0395 & $10^{-1}$ & 8.87 & $\sim T_{\rm struc}$\\
 \hline
  On-res. (max) & 0.1 & 0.1 & 0.00168 & $10^{-2}$ & 0.145  & $\lesssim T_{\rm struc}$ \\ \end{tabular}
 \caption{\label{tab:selfscatter}  Model parameters for the minimal and maximal cases (with respect to the resulting $T_{\rm nt}$)  that are shown in Fig.~\ref{fig:SSrate}. For resonances, self-scattering ceases to be effective at a temperature $T_{\rm struc}$ around the onset of structure formation (the effect of which is not included in Fig.~\ref{fig:SSrate}).
  }
 \end{table}

For illustration, we show in Fig.~\ref{fig:SSrate} the evolution of $\Gamma_{\text{s}}(T)$ and $H(T)$ in the resonant and nonresonant case, for two situations that represent rather extreme cases for the parameter space of the DM model that we consider in Sec.~\ref{sec:models} (see Table \ref{tab:selfscatter} for the relevant model parameters and the resulting  decoupling temperatures). 
As can be seen, the self-scattering rate stays larger than the Hubble rate for a very long time if the Sommerfeld enhancement is resonant and the scattering rate is always able to keep the DM velocity distribution thermal beyond matter-radiation equality at $T_{\text{eq}}\approx 0.75\,$eV \cite{Komatsu:2010fb}. Once matter domination sets in, the Hubble rate $H\propto T^{3/2} g_{*S}^{1/2}(T)$ will catch up with $\Gamma_{\text{s}}$,
 but for many models it stays below the self-scattering rate all the way down to the onset of structure formation (when the formation of gravitational potentials leads to an increase in the DM velocities).  
If we are \emph{not} near a resonance, the intersection of $\Gamma_{\text{s}}$ with $H$ will take place for much larger temperatures:  in our model, we  roughly expect $\mathcal{O}(10)$ MeV $\lesssim T_{\text{nt}} \lesssim \mathcal{O}(100)$ eV. 
 Even when following a conservative approach for the numerical calculation of $\sigma_{T}$ (see Appendix \ref{app:somm_scatt}),  we find  that a thermal velocity distribution is ensured also for all off-resonance models considered here (i.e.~before the annihilations finally come to an end, see below).

In summary, we can safely assume a Maxwellian WIMP velocity distribution in all cases relevant to our discussion -- though we stress that this is a nontrivial statement that needs to be checked explicitly when applying our treatment to other DM models. 
The most noteworthy exception happens if the model parameters  are tuned in such a way as to very efficiently suppress the self-scattering rate due to the {Ramsauer-Townsend effect }\cite{Ramsauer:1921}. 
In this case, the analysis of Eqs.~(\ref{Y_coupled},\ref{y_coupled}) is complicated by the considerably more involved determination of the velocity averages that appear in these equations.
We do not consider this interesting possibility any further here but leave it as a challenge for future studies.

\subsection{Final relic density}
\label{relicdensity}
As explained earlier, a new era of annihilations after chemical (or even kinetic) decoupling is  realized when Sommerfeld enhancements are taken into account. 
We now discuss in some detail three effects that can cause these annihilations to cease eventually.
Obviously, the temperature at which this happens is important for the correct determination of the relic density. 

The first and most important effect is that the Sommerfeld enhancement does not continue to scale like $v^{-1}$ or even $v^{-2}$ as $v\rightarrow 0$ (see Appendix \ref{app:sommerfeld}), but saturates below some cutoff velocity; 
at this point, DM annihilations will no longer be able to keep up with the expansion of the universe. 
 One can estimate the temperature $T_{\text{sat}}$ at which this happens by equating 
 the mean WIMP velocity, $\bar{v} = \sqrt{8 T_{\chi}/ \pi m_{\chi}}$ for a Maxwellian distribution, to 
 $v_{\text{off}}\sim 0.5 \ m_{\phi}/m_{\chi}$ ($v_{\text{on}}\sim \alpha^{3} \ m_{\phi}/m_{\chi}$) for a Sommerfeld enhancement off-resonance (on-resonance), see Eqs.~(\ref{eq:satonres}, \ref{eq:satoffres}).
Assuming that kinetic decoupling takes place more or less instantaneously at $T_{\text{kd}}$, and that we have  $T_{\text{sat}} < T_{\text{kd}}$ (which is the situation we are interested in here),
the saturation temperature can thus be approximated by
\bea
T_{\text{sat,off}} &\sim& m_{\phi} \sqrt{\frac{T_{\text{kd}}}{m_{\chi}}}, \\
T_{\text{sat,on}} &\sim&  \alpha^{3}m_{\phi} \sqrt{\frac{T_{\text{kd}}}{m_{\chi}}}.
\eea

Another effect that is relevant for the discussion of the final relic density is the onset of matter domination. We can see this directly by using
\bea
 H^{2}(T) &=& \frac{4 \pi^{3}}{45 m^{2}_{\text{Pl}}} g_{\text{eff}} T^{4} + \frac{8 \pi}{3 m^{2}_{\text{Pl}}} \rho_{{\rm m},0}\, a^{-3}(T)  \label{eq:Htot}\\
  &\equiv& H_{\text{r}}^2(x)+H_{\text{m}}^2(x)
\eea
in  \eref{deltaY},
where $\rho_{{\rm m},0}= \Omega_{\text{m}} \rho_{\text{c}} \approx 1.10 \times 10^{-47}\text{GeV}^{4}$ is the matter density today \cite{Komatsu:2010fb}. 
Evidently,  $H_{\text{r}}(x) \propto x^{-2}g_{\rm eff}^{1/2}$ and $H_{\text{m}}(x) \propto x^{-3/2} g_{*S}^{1/2}$, i.e.~the effect of matter domination on the evolution of the DM density is essentially the same as replacing $\tilde{n}$ in $\langle\sigma v_{\rm rel} \rangle\propto x^{-\tilde{n}}$ by $n^{\prime}= \tilde{n}+1/2$. As described in Sec.~\ref{reenter_annihilation}, an appreciable change in $Y$ is then only possible for $n^{\prime} \le -1$, which is now no longer achieved with a $1/v$ enhancement after kinetic decoupling ($n^{\prime}=-1/2$). Only in the special case that we are on a resonance, we would have $n^{\prime}=-3/2$ after kinetic decoupling has taken place, and annihilations could continue even beyond matter-radiation equality.

The last effect, which  eventually brings all cosmological WIMP annihilations to an end, is the onset of structure formation around $z_{\rm struc}\sim \mathcal{O}(100)$: once significant gravitational potentials are formed, they start to attract the DM particles and cause their average  velocity to increase again, eliminating the Sommerfeld effect.

For the models that we are interested in here, to be introduced in Sec.~\ref{sec:models}, we  always find
\be
  T_{\text{sat,off}} > T_{\text{nt}} \gg T_{\text{eq}}\,.
\ee
This means that the final end of DM annihilation is triggered by the saturation of the Sommerfeld effect at small velocities as long as the particular combination of DM and exchange particle masses do not put us on a resonance.
On a resonance, on the other hand, the saturation of the Sommerfeld effect happens much later and we find cases in which the WIMPs  continue to annihilate until well after matter-radiation equality (though we always have $ T_{\text{sat,on}} > T_{\text{struc}} \gtrsim T_{\text{nt}}$).

Taking into account the appropriate saturation effect(s) described above, let us now denote with  $Y_0$ the result of integrating \eref{Y_coupled} to $x_0=m_\chi/T_0$, where $T_0=2.348 \times 10^{-4}$ eV is the photon temperature of the universe today. The  DM relic density, in units of the critical density,
is then obtained as \cite{Nakamura:2010zzi}
\be
  \Omega_\chi = m_\chi s_0 Y_0/\rho_c=2.742 \times 10^{11} h^{-2} \left(\frac{m_\chi}{\rm TeV}\right) Y_0\,.
\ee
For viable DM models, this should be compared to 
\be
   \Omega_{\rm DM}=0.229\pm0.015\,,
\ee
which is the most recent estimate from observations \cite{Komatsu:2010fb}. If the DM particle is its own antiparticle, $\chi=\bar\chi$, we should thus demand $\Omega_\chi= \Omega_{\rm DM}$; otherwise we have $\Omega_\chi= \Omega_{\rm DM}/2$.

\subsection{The smallest protohalos}
\label{protohalos}

A new era of annihilations after kinetic decoupling not only affects the DM relic density, but also the DM velocity dispersion at the onset of structure formation. This quantity, in turn, can be related to a small-scale cutoff  $M_{\rm cut}$ in the power spectrum of matter density fluctuations which corresponds to the mass of the smallest gravitationally bound objects \cite{Green:2005fa}. 

Let us therefore use the asymptotic value of $y$, when both DM  scattering and annihilation events have finally come to an end to define an \emph{effective asymptotic decoupling temperature}  
\be
  x_{\rm dec}^\infty=\frac{m_\chi}{T_{\rm dec}^\infty}\equiv \left.y\right|_{x\rightarrow\infty}
  \times\left.\frac{s^{2/3}}{T^2}\right|_{T=T_{\rm dec}^\infty} \,.
\ee
Note that this temperature can be quite a bit higher than the temperature at which the annihilations actually stop. It thus does not have any obvious intuitive interpretation (unless there is no second era of annihilation, in which case it is simply given by $T_{\rm dec}^\infty=T_{\rm kd}$), but is only introduced here for convenience because it allows to express the asymptotic DM temperature, also known as velocity dispersion, as $T_\chi=T^2/T_{\rm dec}^\infty$.

As a consequence, we may simply replace $T_{\rm kd}\rightarrow T_{\rm dec}^\infty$ in any expression that relates the kinetic decoupling temperature to the cutoff in the power spectrum of matter density fluctuations. In particular, we have \cite{Bringmann:2009vf}
\be
  M_{\rm cut}=\max\left[M_{\rm fs},M_{\rm ao}\right]\,,
\ee
where free-streaming  \cite{Hofmann:2001bi,Green:2005fa} induces an exponential cut-off characterized by
\be
 \label{mfs}
   M_{\rm fs}\approx2.9\times 10^{-6}\left(\frac{1+{\rm ln}\left(g_{\rm eff}^{1/4}T_{\rm dec}^\infty/50\;{\rm MeV}\right)/19.1}{\left(\frac{m_\chi}{100\; {\rm GeV}}\right)^{1/2} g_{\rm eff}^{1/4}\left(\frac{T_{\rm dec}^\infty}{50\;{\rm MeV}}\right)^{1/2}}\right)^3M_\odot\,.
\ee
and acoustic oscillations \cite{Loeb:2005pm,Bertschinger:2006nq}  a similar  cutoff with 
\be
 \label{mao}
  M_{\rm ao}\approx3.4\times10^{-6}\left(\frac{T_{\rm dec}^\infty g_{\rm eff}^{1/4}}{50\,{\rm MeV}}\right)^{-3}M_\odot\,.
\ee

We note that the above  prescription for calculating $M_{\rm cut}$ can, strictly speaking, only be applied if $f$ (i) reaches its asymptotic behavior  (given by $y\simeq  \left.y\right|_{x\rightarrow\infty}$) already well before the onset of matter domination and (ii) has a velocity distribution at that time which resembles that of the standard WIMP case -- simply because these are the assumptions that entered into the analysis of Refs.~\cite{Hofmann:2001bi,Green:2005fa,Loeb:2005pm,Bertschinger:2006nq}. As we will see, 
it is possible to find viable models where these assumptions are not satisfied; in these cases, our prescription thus only gives an approximative value for $M_{\rm cut}$. While we do not pursue this issue any further here, it would certainly make for an interesting study to extend the standard analysis and investigate how the spectrum of matter density fluctuations would be  affected in such cases.

\section{Leptophilic DM models}
\label{sec:models}
In this Section, we demonstrate how to apply our general discussion to a specific class of WIMP models that exhibit large Sommerfeld effects.
As already mentioned in the introduction, the cosmic ray  lepton excess \cite{CR_excess} has triggered a lot of phenomenological activity, trying to establish a possible DM connection. At first sight, this is not a trivial task at all since standard WIMP candidates, like the neutralino,  fail to meet the necessary criteria from a model-independent analysis of the data \cite{Cirelli:2008pk,Bergstrom:2009fa} in order to fit the observed excess,  the annihilation of DM particles would have to (i) produce a very hard positron spectrum at least up to a few hundred GeV, but (ii) at the same time  produce almost no antiprotons, and iii) happen with a rate about 2 to 3 orders of magnitude above the canonical value of $3\cdot10^{-26}{\rm cm}^3{\rm s}^{-1}$ which is needed for thermal production in the simplest scenarios (see, however, Ref.~\cite{Bergstrom:2008gr} for a way of how neutralino DM could at least satisfy the two first requirements).

An elegant, or at least very economical, way to meet all these requirements is to postulate the existence of a new light exchange boson $\phi$, with $100\,{\rm MeV}\lesssim m_\phi\lesssim1\,$GeV, that directly couples to a DM particle with $m_\chi\sim\mathcal{O}(1\,{\rm TeV})$  but only very weakly to standard model particles (in order to avoid the stringent bounds on light new particles). The DM annihilation rate today would then be strongly enhanced by the Sommerfeld effect, without changing too much the annihilation rate during freeze-out; furthermore, the decay of the resulting $\phi$ particles into hadronic modes is kinematically forbidden, thus leaving only the desired light leptonic modes (see Ref.~\cite{ArkaniHamed:2008qn} for a general account of this idea). 

Many realizations and variations of this rather general setup have been worked out in, e.g., Refs.~\cite{Pospelov:2008jd,Cholis:2008qq,Nomura:2008ru,Fox:2008kb}. Noting that an astrophysical explanation of the above-mentioned cosmic ray excess may well be more likely \cite{Serpico:2011wg}, we will here only consider a very simple phenomenological toy-model for leptophilic WIMP DM -- loosely motivated by the cosmic ray lepton data but mostly chosen for illustration of  the possible effects of the Sommerfeld enhancement on the thermal decoupling process. 

\subsection{A simple toy-model}
\label{sec:toymod}

We consider a fermionic DM particle $\chi$ that couples only to a light scalar $\phi_s$ and a pseudoscalar $\phi_p$ via
\be
  \mathcal{L}\supset g^s_\chi\phi_s\bar\chi\chi+g^p_\chi\phi_p\bar\chi\gamma^5\chi\,.
\ee
We assume that the (pseudo-)scalar particles also interact with standard model leptons, through
\be
  \mathcal{L}\supset g^s_\ell\phi_s\bar\ell\ell+g^p_\ell\phi_p\bar\ell\gamma^5\ell\,,
\ee
albeit with much smaller coupling strengths (i.e.~$g^s_\ell$ and $g^s_\ell$ can be thought of as effective couplings arising, e.g., from higher-dimension operators). 

For simplicity, we will assume $ g^{s,p}_e=g^{s,p}_\mu=g^{s,p}_\tau$. Possible couplings to quarks typically do not change the phenomenology of our model (for the mass ranges considered here) and are anyway strongly constrained -- see Appendix \ref{app:Constraints} for experimental bounds on this and similar models; including nonzero neutrino couplings would not have a large impact on the results either, since the DM scattering off neutrinos is negligible due to the small neutrino masses $m_{\nu}$, see also \eref{chifchif}. The scalar particle mediates DM scattering at low-momentum transfer and is also responsible for the Sommerfeld enhancement. The need for an additional pseudoscalar particle arises because of parity conservation: While the annihilation $\chi\chi\rightarrow\phi_s\phi_p$ has an $s$-wave contribution, $\chi\chi\rightarrow\phi_s\phi_s (\phi_p\phi_p)$ is a $p$-wave process which vanishes in the $v\rightarrow0$ limit, so it is only the former  channel that can happen with a sizable rate today (which, in turn, is required if we want to make any contact to observable cosmic ray lepton fluxes at all). Even the relic density is to a large extent  determined through this channel and thus mainly depends on the parameter
\be
 \label{defalpha}
\alpha\equiv\frac{g^s_\chi g^p_\chi}{4\pi}\,.
\ee
Since we want to keep the discussion simple, we will mostly assume $g^s_\chi= g^p_\chi$ in the following, noting that the relic density is essentially only set by \emph{one} effective coupling constant anyway.

We note that our setup  roughly corresponds  to a simple version of the model proposed by Nomura and Thaler \cite{Nomura:2008ru}, where the new particles postulated above $(\chi, \phi_s, \phi_p)$ are embedded in a full supersymmetric scenario and $\phi_p$, in particular, takes the role of an almost ``standard'' Peccei-Quinn axion. In Appendix \ref{app:m2}, we collect the relevant annihilation cross-sections and scattering matrix elements for this model.

\begin{figure}[t]
	\includegraphics[width=\columnwidth]{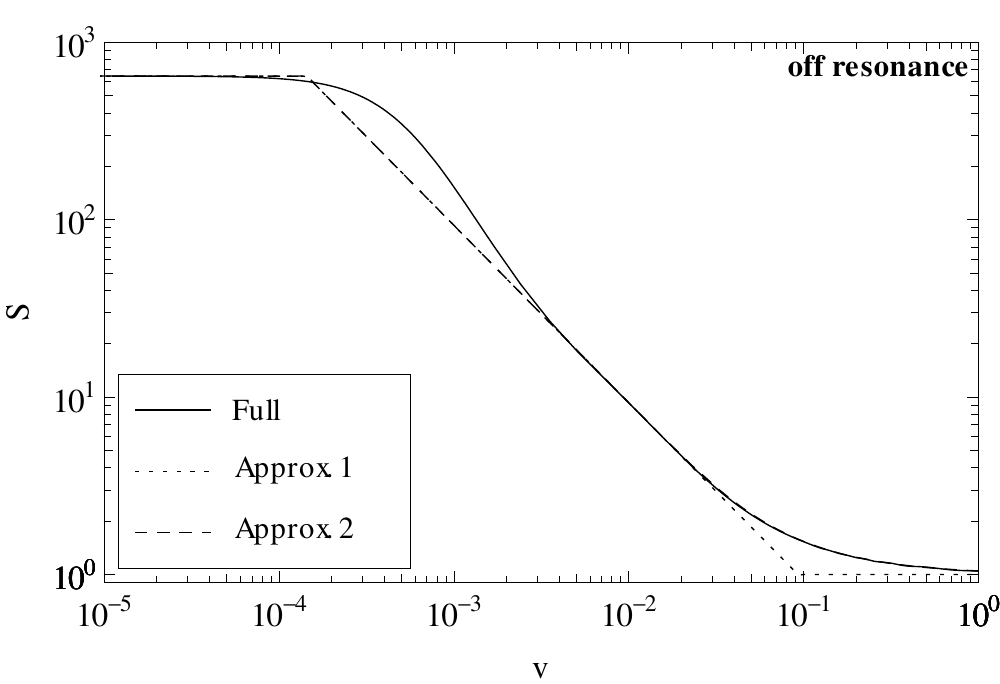}
	\caption{Shown are two approximations (dotted, dashed) to the numerically calculated, full Sommerfeld enhancement (solid), for $m_{\chi}=1$ TeV, $m_{\phi}=5$ GeV and $\alpha=0.03$.
See text for further details. \label{fig:SE_offres}} 
\end{figure}

\subsection{Decoupling on- and off-resonance}

\begin{figure*}[t]	
	\includegraphics[height=0.67\columnwidth]{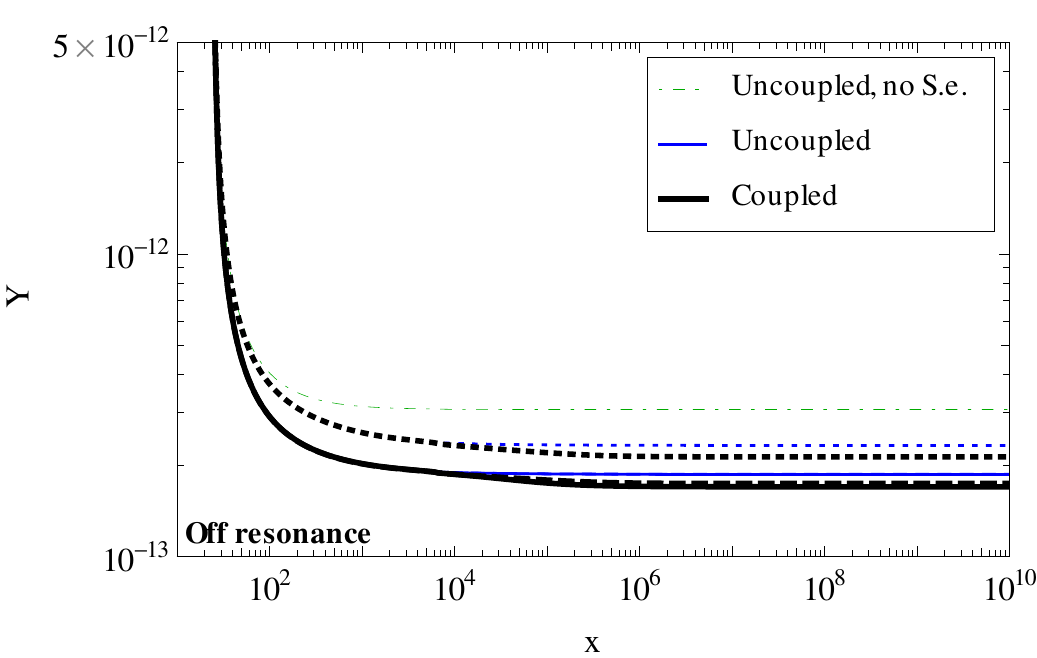}
	\includegraphics[height=0.67\columnwidth]{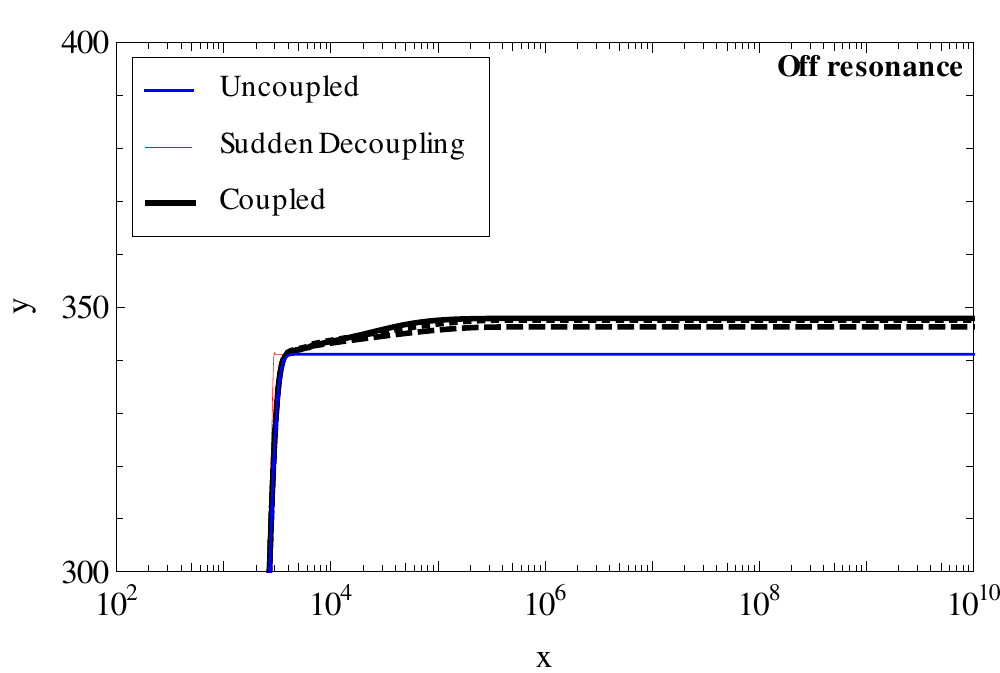}
	\caption{For a parameter set where the Sommerfeld enhancement is not near a resonance ($m_{\chi}=1$ TeV, $m_{\phi}=5$ GeV, $\alpha=0.03,$
 and $g_{\ell} = 10^{-7}$), we show the evolution of the quantities $Y$ as defined in \eref{Ydef}, and $y$ as defined in \eref{ydef}. We show the solution to the full set of coupled Boltzmann equations (black) as well as for two approximations described in more detail in the text, assuming sudden kinetic decoupling (red; not visible in left-hand plot) and no coupling between the Boltzmann equations for $y$ and $Y$ (blue), respectively. Different approximations to the Sommerfeld enhancement are shown according to Fig.~\ref{fig:SE_offres} by dotted, dashed, or solid lines. Additionally, the  solution without Sommerfeld enhancement is shown in the left-hand panel (green, dash-dotted). \label{fig:offres}}
\end{figure*}

For this model, let us first study  in some detail the evolution of the WIMP number density and temperature for two cases of particular interest, i.e.~parameter sets for which we are outside and exactly on a resonance, respectively. 
We will be especially interested in quantifying the difference between our full treatment, described in Secs.~\ref{sec:decoupling} and \ref{sec:new_ann}, and various approximations one may deem reasonable. In particular, we will refer to the \emph{full} solution as the coupled set of Eqs.~(\ref{dYdx}, \ref{dydx}), with the proper replacement of $T\rightarrow T_\chi$ in the expressions for $\langle\sigma v_{\rm rel}\rangle_{(2)}$; for $Y\gg Y_{\rm eq}$, this simply corresponds to Eqs.~(\ref{Y_coupled}, \ref{y_coupled}). We denote with \emph{sudden decoupling} the case in which we assume $T_\chi(T>T_{\rm kd})=T$ and $T_\chi(T\leq T_{\rm kd})=T^2/T_{\rm kd}$ in \eref{dYdx} and neglect the term proportional to $Y'$ in \eref{dydx}. The \emph{uncoupled} solution, finally, corresponds to the case where kinetic decoupling is assumed to have no influence on the evolution of the WIMP number density, and vice versa.

In addition to the full numerical solution for the Sommerfeld factor, we also considered two different approximations as illustrated in Fig.~\ref{fig:SE_offres}. For approximation $1$ (shown as a dotted line) we assumed $S=1$ for $v>\pi \alpha$, $S=\pi \alpha/ v$ in the intermediate regime and $S(v\leq v_{\rm max})\equiv S_{\rm max}$ (defined by the full solution). As a better approximation, especially at larger velocities, we used the Coulomb approximation $S=\frac{\pi \alpha}{v}/(1-e^{-\frac{\pi \alpha}{v}})$ down to $v_{\rm max}$, and $S=S_{\rm max}$ for smaller velocities. This approximation $2$ is shown as a dashed line. 
For the case of resonances, we adjusted these approximations correspondingly: in approximation $1$ we used  $S\propto v^{-2}$ instead of $S\propto v^{-1}$, and in approximation $2$ we used the Coulomb expression down to velocities where the enhancement follows the $v^{-2}$ behavior.

In Fig.~\ref{fig:offres},  the solutions for $y$ and $Y$ are shown for one particular parameter set for which the Sommerfeld enhancement is not in the neighborhood of a resonance. As expected, the relic density is lower (by 10\%) for the coupled solution than for the uncoupled solution; compared to the case without Sommerfeld enhancement, it is smaller by $80\, \%$. Also visible is that approximation $2$ reproduces the full numerical result much better than approximation $1$, for which the relic density is overestimated. Assuming that kinetic decoupling happens suddenly, like depicted in the right-hand panel of Fig.~\ref{fig:offres} (shown in red), gives a reasonable approximation to the full evolution of $y$ -- though we find that for very small lepton couplings, and negligible $\phi$-scattering, the kinetic decoupling process may be more delayed than for standard WIMPs \cite{Bringmann:2006mu, Bringmann:2009vf}: since more scattering events are necessary to keep the DM in thermal equilibrium in this case, it also takes longer before  scattering becomes sufficiently inefficient so that $y$ remains completely constant.  
Therefore, sudden decoupling works very well as an approximation to $Y$ (the lines are completely obscured by the full solution in Fig.~\ref{fig:offres}, right). For this set of parameters, kinetic decoupling happens at $x_{\rm kd}\sim2.92 \times 10^{3}$, after which the coupled solutions start to deviate from the uncoupled solutions. The annihilations cease around $x\sim3\times10^{5}$ because of the saturation of the Sommerfeld factor, and both $y$ and $Y$ stay constant afterwards.

\begin{figure*}[t]
	\includegraphics[width=\columnwidth]{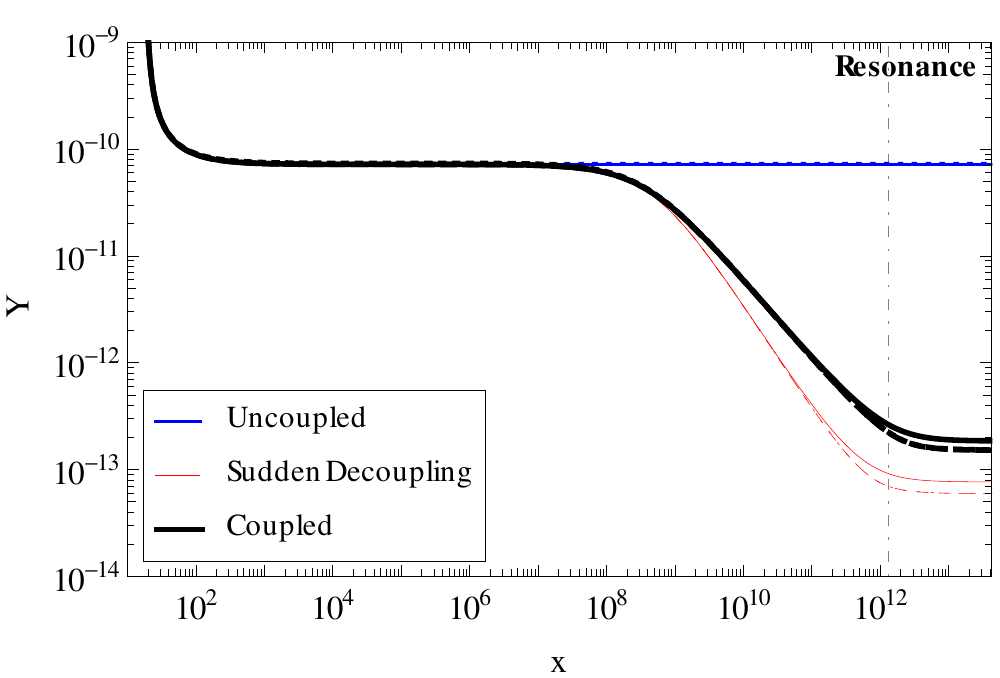}
	\includegraphics[width=\columnwidth]{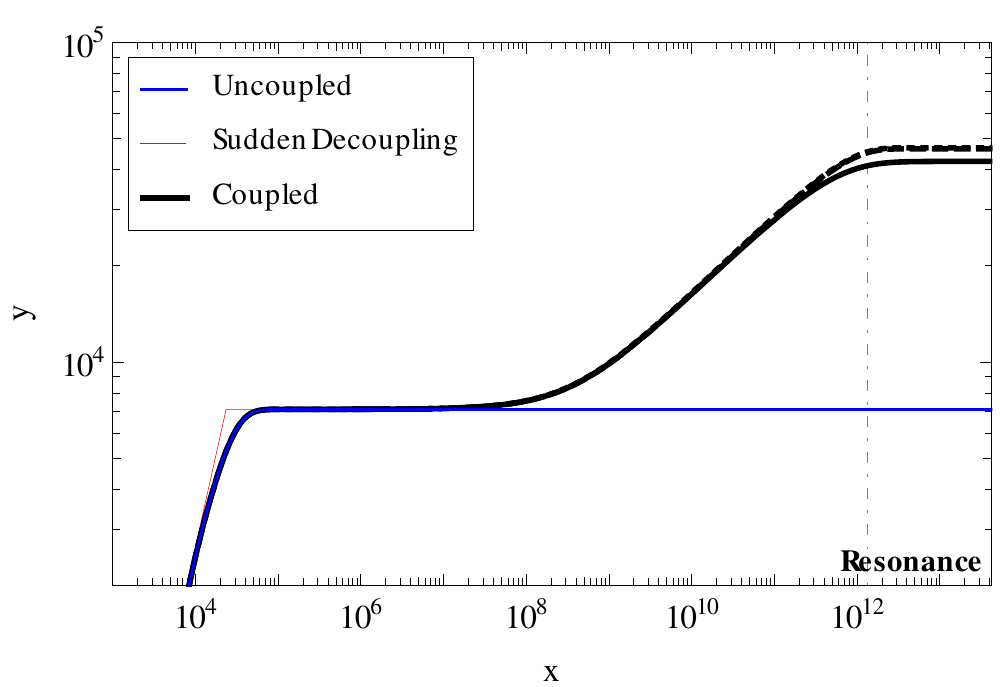}
	\caption{Same as Fig.~\ref{fig:offres}, but now for a parameter set where the Sommerfeld enhancement is resonant ($m_{\chi}=1$ TeV, $m_{\phi}=1$ GeV, $\alpha=0.00168$ and $g_{\ell} = 4.6 \times10^{-5}$). It can be seen that annihilations continue even until after matter-radiation equality (denoted as a dash-dotted line). }\label{fig:res}
\end{figure*}

The more interesting case is to consider a parameter set for which we are on a resonance in the annihilation cross section. The results, shown in Fig.~\ref{fig:res}, show a striking difference with respect to the uncoupled Boltzmann equations for $x\gtrsim 10^{7}$, after which the WIMP annihilations decrease the relic density by more than two orders of magnitude until the Sommerfeld enhancement saturates at $x\simeq 2\times10^{13}$. Following the argument presented in Sec.~\ref{reenter_annihilation}, one might have expected an even more efficient decrease in the relic density that  starts directly after kinetic decoupling (which for this choice of parameters happens at $x_{\rm kd}\simeq2.35 \times10^{4}$). This is not observed, however, because at early times the velocities of the WIMPs are still  large enough to be in the Coulomb regime and, even though we are on a resonance, the Sommerfeld enhancement follows an $S\propto1/v$  rather than $S\propto 1/v^{2}$ behavior (see Appendix \ref{app:sommerfeld}). Indeed, for $3 \times 10^{5}<x\lesssim 3 \times10^{6}$ we find that \eref{eq:asymprelationyY} is satisfied with $-2<\tilde{n}\leq-1$, whereas the Sommerfeld enhancement shows the resonant behavior ($\tilde{n}=-2$) for larger values of $x$. 

The value for the coupling $\alpha$ is here chosen such that the full solution gives the right relic density today (within 3$\sigma$): the uncoupled solution (both with and without including the full Sommerfeld factor) actually overestimates $Y_{0}$ by a factor of $\sim 400$. Sudden decoupling becomes a very bad approximation to $y$ in the case of resonances, at least for $x\gtrsim10^{7}$ where it simply follows the uncoupled solution. However, it catches the overall behavior of a large decrease in $Y$ rather well -- though it underestimates the final relic density by at least a factor of $2$. The different approximations to the Sommerfeld enhancement, on the other hand, give comparable results.

\subsection{Range of decoupling temperatures and the mass of the smallest protohalos}
\label{sec:range}

\begin{figure}[t!]
	\includegraphics[width=\columnwidth]{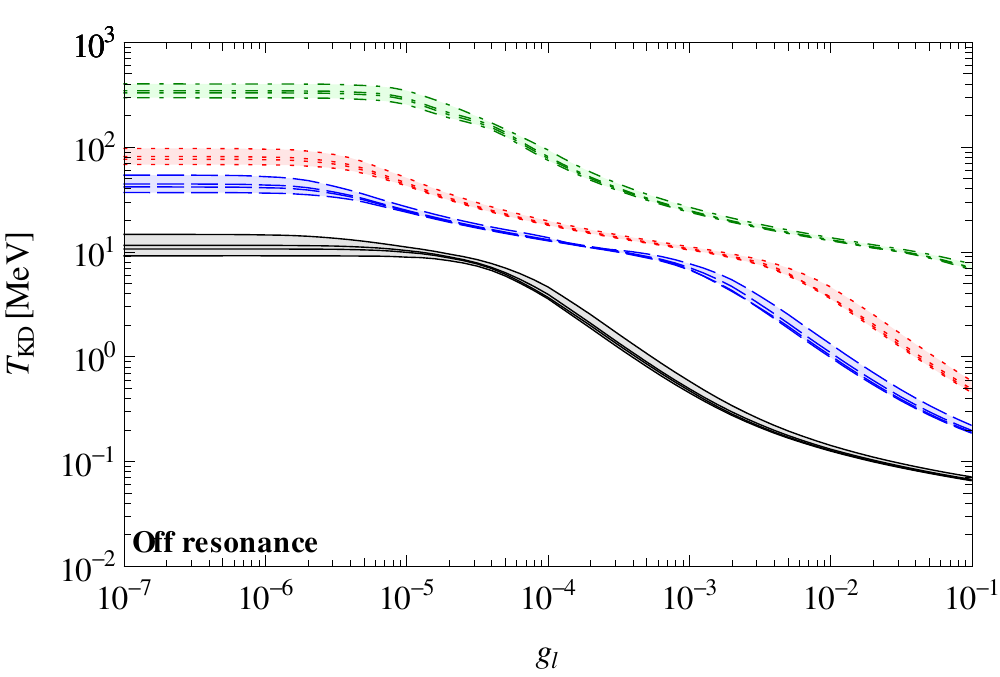}
	\caption{For models where the Sommerfeld enhancement is not on a resonance, the kinetic decoupling temperature is shown as a function of the mediator particle coupling to leptons, for $m_{\phi} = 100$ MeV (black, full), $500$ MeV (blue, dashed), $1$ GeV (red, dotted), and $5$ GeV (green, dash-dotted). From bottom to top, the lines correspond in each case to a DM mass of $m_{\chi}=100, 500, 1000, 5000$ GeV.  \label{fig:tkdoffres}}
\end{figure}

Having discussed in some detail the situation for two particular parameter sets, 
let us now explore the thermal history of DM, and, in particular, the consequences of our improved treatment, for the full possible range of our model parameters. 
For this purpose,  we adjusted the  coupling $\alpha$ in all calculations in such a way that the relic density obtained by solving the full, coupled Boltzmann equations is within $3 \sigma$ of the observed value today, $0.184 \leq \Omega_{\rm DM} \leq 0.274$, leaving thus the lepton coupling $g_{\ell}$ as the only other free parameter besides $m_{\chi}$ and $m_{\phi}$.

Let us start by showing in  Fig.~\ref{fig:tkdoffres} the kinetic decoupling temperature $T_{\rm kd}$ as a function of the coupling constant $g_{\ell}$. As can be seen, $T_{\rm kd}$ decreases for larger $g_{\ell}$ -- reflecting the fact that a strong lepton coupling will keep the WIMPs longer in local thermal equilibrium. For decoupling temperatures smaller than around 7\,MeV, only DM scattering with electrons is effective just before the DM particles completely leave thermal equilibrium. 
At higher temperatures, also muons start to contribute very efficiently to the scattering process, resulting in a flattening of  $T_{\rm kd}(g_{\ell})$ when moving to smaller values of $g_{\ell}$; this already happens for highly nonrelativistic muons because of the $m_\ell^2$ dependence of the scattering rate, see \eref{chifchif}.
Scattering with mediator particles ensures that the kinetic decoupling temperature does not increase arbitrarily high even for negligible lepton couplings, which explains the plateau that appears at roughly $g_{\ell}\lesssim10^{-5}$. The decoupling temperature $T_{\rm kd}$ increases rather strongly  with higher mediator masses $m_\phi$: in the case of lepton scattering this directly follows from the form of the scattering matrix element, \eref{chifchif}, while in the case of scattering with the mediator particles it reflects the strong Boltzmann suppression of the latter.
The dependence of the decoupling temperature on the DM mass  $m_\chi$, on the other hand, is very weak. Also the spread in $T_{\rm kd}$ (for given values of $m_{\phi}, m_{\chi}$, and $g_{\ell}$) due to the different values of the relic density that were obtained by changing the coupling $\alpha$ accordingly is essentially negligible. 

\begin{figure}[t!]
	\includegraphics[width=\columnwidth]{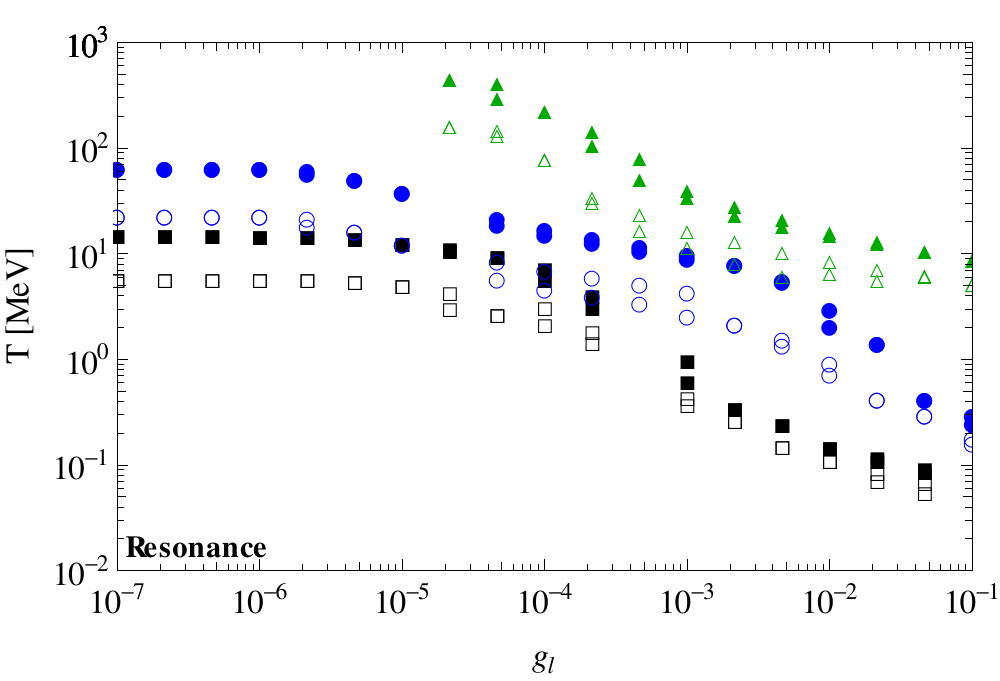}
	\caption{For models where the Sommerfeld enhancement is resonant, the kinetic decoupling temperature $T_{\rm kd}$ (filled) in comparison to the effective asymptotic decoupling temperature $T^{\infty}_{\rm dec}$(empty) is shown. Again $m_{\phi} = 100$ MeV (black, squares), $500$ MeV (blue, circles), and $5$ GeV (green, triangles). \label{fig:tkdmcutonres}}
\end{figure}

In Fig.~\ref{fig:tkdoffres} we have chosen to only show models with parameter sets where the Sommerfeld enhancement is not resonant both because this is in some sense the generic behavior of the model and because the parameter dependence is very straightforward to discuss in that case.
Phenomenologically, on the other hand, the resonant case is also very interesting, not the least because much stronger effects compared to the standard scenario can be expected. We therefore performed a dedicated scan to sample that part of the parameter space that results in a resonant Sommerfeld enhancement \emph{and} the correct relic density (in practice, we only sampled the first 5 resonances; higher resonances are more and more densely distributed, increasing thus the amount of parameter fine-tuning that is necessary to meet the just mentioned criteria).
As shown in Fig.~\ref{fig:tkdmcutonres}, the resulting kinetic decoupling temperature is a bit higher than in the offresonance case. This is due to the fact that the relic density constraint makes a smaller coupling $\alpha$  necessary to compensate for the large Sommerfeld effect, which in turn decreases the amount of DM scattering off $\phi_{s}$ so that WIMPs decouple slightly earlier than in the off resonance case. The  \emph{asymptotic} decoupling temperature  $T^{\infty}_{\rm dec}$, on the other hand, is up to a factor of $\sim 5$ smaller than $T_{\rm kd}$ and always smaller than in the off-resonant case (note that off-resonance $T^{\infty}_{\rm dec}$ and $T_{\rm kd}$ differ, as expected, by only 3\,\% at most).

\begin{figure}[t]
	\includegraphics[width=\columnwidth]{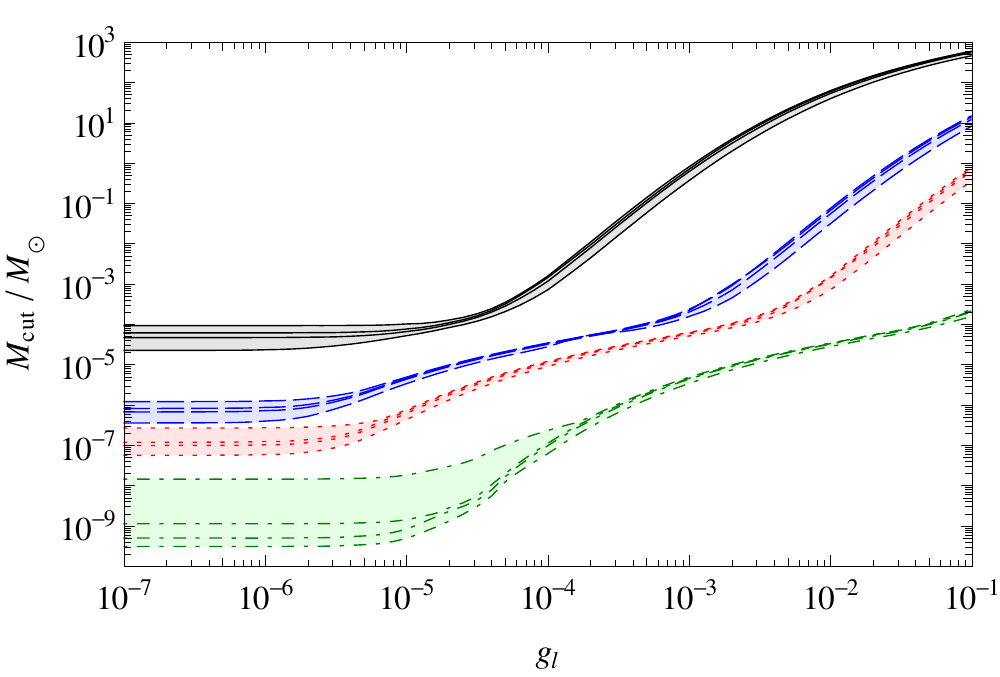}
	\caption{For models where the Sommerfeld enhancement is not on a resonance, the cutoff mass is shown as a function of the lepton coupling, for $m_{\phi} = 100$ MeV (black, full), $500$ MeV (blue, dashed), $1$ GeV (red, dotted), and $5$ GeV (green, dash-dotted). From top to bottom, the lines correspond in each case to a DM mass of  $m_{\chi}=100, 500, 1000, 5000$ GeV.  On-resonance, the value of $M_{\rm cut}$ can be larger by a factor of up to 2 (20) for very large (small) values of $g_\ell$.\label{fig:mcutoffres}}
\end{figure}

From the asymptotic decoupling temperature, we also calculated the corresponding mass of the smallest gravitationally bound objects; the off-resonance results are shown in  Fig.~\ref{fig:mcutoffres}. It can be seen that the possible cutoff mass spans a wide range of $M_{\rm cut}/M_\odot \sim \mathcal{O}(10^{-10}$ -- $10^{3})$. Even when taking into account existing constraints on the lepton coupling, see Appendix \ref{app:Constraints},  much larger cutoff masses than in the standard WIMP case thus seem possible; imposing $g_\ell<10^{-3}$, e.g.~results in $M_{\rm cut} \lesssim M_\odot$ (whereas $M_{\rm cut} \lesssim 10^{-3} M_\odot$ for neutralino DM \cite{Bringmann:2009vf}).
Compared to $T_{\rm kd}$ and $T_{\rm dec}^{\infty}$, $M_{\rm cut}$ can have a stronger dependence on $m_{\chi}$, visible by the spread of the bands for $g_{\ell}\lesssim 10^{-4}$. This is due to the fact that in this region of parameter space the free-streaming mass $M_{\rm fs}$ (that depends, among other things, on $m_{\chi}$), dominates over the acoustic oscillation cutoff mass $M_{\rm ao}$. In the resonant case, the resulting $M_{\rm cut}$ is in general higher; naively using $T_{\rm kd}$ instead of $T_{\rm dec}^{\infty}$ to calculate it would result in values up to 2 orders of magnitude smaller. As expected from Fig.~\ref{fig:tkdmcutonres}, the effect of resonances is most pronounced for small values of $M_{\rm cut}$, corresponding to large values of $T_{\rm kd}$;
as a result, the \emph{lowest} possible cutoff mass shifts from $\sim3 \times 10^{-10} M_\odot$ to  $\sim7 \times 10^{-9} M_\odot$, while the largest possible value shifts from $\sim600  M_\odot$ to $\sim1100  M_\odot$.

Let us now discuss the evolution of the WIMP number density and the resulting relic density.
Annihilations off-resonance finally come to an end for temperatures ranging from $T_{\rm sat}\sim 100$ MeV to $\sim1$ keV, where $T_{\rm sat}$ was defined by $Y(T_{\rm sat})/ Y_0\equiv 0.99 $. Whereas the maximum saturation temperature is more or less independent of $m_{\phi}$, the minimum $T_{\rm sat}$ is significantly smaller for small mediator masses. This comes from the fact that the Sommerfeld enhancement increases for smaller values of $m_{\phi}/(m_{\chi} \alpha)$, so the new era of annihilations after kinetic decoupling can last longer for smaller $m_{\phi}$; furthermore, kinetic decoupling happens later in this case. 
We also find that, as expected, $T_{\rm sat}$ decreases with decreasing $g_{\ell}$, i.e.~with increasing $T_{\rm kd}$. Finally, we observe that for many models with large lepton couplings we have $T_{\rm sat}>T_{\rm kd}$, indicating that the effect of reentering an era of annihilations is negligible, whereas for small $g_{\ell}$ we find that $T_{\rm kd}/T_{\rm sat}$ can reach values up to $\mathcal{O}(10^{4})$. 

For nonresonant Sommerfeld enhancement, the resulting effect of the new era of annihilations on $Y$ is significant but never extremely large: the ratio of the relic density obtained by the uncoupled Boltzmann equations with that obtained by the full solution is at most $\Omega_{\chi\rm ,u}/\Omega_{\chi,\rm c}\lesssim 1.1$ for the range of parameters that we have scanned. This ratio tends to be slightly larger for higher WIMP masses, reflecting the fact that a larger Sommerfeld effect due to an increased value of $\alpha$ is expected in that case (recall that the relic density is set by the annihilation rate which scales, roughly,  as $\left<\sigma v\right> \propto \alpha^{2}/ m^{2}_{\chi}$). We also find that, as expected, $\Omega_{\chi\rm ,u}/\Omega_{\chi,\rm c}$  grows with increasing $T_{\rm kd}$: the new era of annihilations simply lasts longer when kinetic decoupling takes place early.

\begin{table}[t!]
 \centering
 \begin{tabular}{l||c|c}
    &  Off-resonance  & On-resonance \\
  \hline
  \hline
  $x_{\rm cd}$ & $\sim24$ -- 27 & $\sim20$ -- 25\\
  \hline
  $T_{\rm kd}$[MeV] & $\sim0.07$ -- 400 & $\sim0.09$ -- 450 \\
  \hline
  $T_{\rm dec}^{\infty}$ [MeV] & $\sim0.07$ -- 400 & $\sim0.06$ -- 170\\
  \hline
  $T_{\rm sat}$ [keV] & $\sim1$ -- $10^{5}$& $\sim10^{-6}$ -- $10^{-2}$ \\
  \hline
   $\Omega_{\chi, \rm u}/\Omega_{\chi, \rm c}$ & $\sim1$ -- 1.1 & $\sim3.5$ -- 670\\
   \hline
  $M_{\rm cut} [M_{\odot}]$ &$\sim 3 \times 10^{-10}$ -- 600 & $ \sim 7 \times 10^{-9}$ -- 1100\\
 \end{tabular}
 \caption{\label{tab:offonres} Overview of the various decoupling temperatures in our toy-model, for the parameter range that  we considered here. Also stated is the resulting change in the relic density when fully taking into account the coupled Boltzmann equations [Eqs.~(\ref{dYdx}, \ref{dydx})] rather than ignoring the possible impact of kinetic decoupling; finally, we summarize the possible range of cutoff masses (see also text and Fig.~\ref{fig:mcutoffres}).
  }
 \end{table}

In the resonant case, the change in $Y$ by considering the coupled Boltzmann equations rather than  the uncoupled equation is much more significant, yielding a relic density that can be up to a factor of $\sim$670 smaller. In most cases annihilations even continue after matter-radiation-equality, further decreasing the relic density in some extreme cases by up to a factor of $4$ before reaching its final value: $\Omega_{\chi}(T_{\rm eq}) \lesssim 4 \, \Omega_{\chi} (T_{0})$. 
The relic density correspondingly saturates at very low temperatures, $\mathcal{O}(10)$ eV $\gtrsim T_{\rm sat}\gtrsim \mathcal{O}(10^{-3})$ eV. 
Since the annihilations are more efficiently suppressed after the onset of matter domination, the relic density could probably not continue to decrease a lot more, also because of the already mentioned increase in the WIMP velocity  once gravitational potentials build up and cosmological structure formation starts. The effect we see here due to resonances is therefore (close to) the maximum effect on the relic density that can be expected, and increasing the Sommerfeld enhancement (e.g.~by an even denser sampling of the model parameters near the resonances) are unlikely to have a large impact.

For convenience, we summarize in Table~\ref{tab:offonres} the above discussion by showing the range of possible decoupling temperatures that we have encountered in our scan of model parameters, along with the resulting change in the relic density and possible values for the mass of the smallest protohalos.

\section{Discussion}
\label{sec:disc}

In this Section, we return to some technical issues that we have not addressed explicitly in Sec. \ref{sec:models}, indicate possible extensions to our analysis and discuss possible consequences of our findings. 

One such comment concerns the scattering of DM with mediator particles which we have tacitly assumed to be in thermal equlibrium in the above analysis: we actually checked  that this is always satisfied for $T\gtrsim T_{\rm kd}$ because of the very efficient (inverse) decay processes $\phi\leftrightarrow \ell^+\ell^-$ (for more details, see Appendix \ref{app:m2}). If one would neglect $\chi\phi\leftrightarrow\chi\phi$ scatterings, and only take into account DM scattering with standard model leptons, the kinetic decoupling temperature would continue to increase without any bounds for smaller values of $g_\ell^\phi$. As an interesting consequence, it is thus only due to the presence of a thermal population of $\phi$ particles that a situation with $T_{\rm kd} > T_{\rm cd}$ cannot occur (note also that the thermal production mechanism for $\chi$ actually \emph{requires} the exchange particles to be in thermal equilibrium around $T_{\rm cd}$; 
this implies a \emph{lower} bound on $g_\ell^\phi$ such that one essentially cannot get around this argument even for a much wider class of DM models than considered here). 

Let us also mention that in Sec.~\ref{sec:models}, for simplicity, we only took into account DM  scattering with \emph{scalar} particles. Including also the scattering with pseudoscalars affects only the kinetic decoupling temperature directly: other observables, such as the relic density, change indirectly because of the coupling between $Y$ and $y$. In this case, we find that $T_{\rm kd}$ for $g_{\ell} \lesssim 10^{-4}$ becomes as expected a bit lower (at most $\sim20 \%$ for $g_\ell^s=g_\ell^p$ and $m_s=m_p$), with a correspondingly smaller impact on, e.g., the relic density.

Furthermore, we have in our discussion always assumed that the DM particles couple with the same strength to both scalars and pseudoscalars, i.e.~$g^{s}_{\chi}=g^{p}_{\chi}$. 
In principle, however, the coupling strength $\alpha$, cf.~Equation \eqref{defalpha}, that roughly sets the relic density is not the same as the coupling strength $\alpha^s\equiv {g^s_\chi}^2/{4\pi}$ that governs the Sommerfeld enhancement. Allowing for $g^s_\chi\neq g^p_\chi$ would thus imply a larger range of possible $\alpha^s$ that is consistent with the relic density requirement. While we choose  not to explore the full phenomenology of this option here, we note that the implications for the off-resonance case are not expected to be sizable because a relatively large value of $\alpha$ severely restricts the range of $\alpha^s$ (neither $g^s_\chi$ nor $g^p_\chi$ must be too large in order to remain in the perturbative regime); if $g^s_\chi$ and $g^p_\chi$ are varied independently, on the other hand, less fine-tuning is in some sense required to arrange for a resonant Sommerfeld enhancement.
In order to make the first remark a bit more quantitative, let us consider the situation of keeping $\alpha$  fixed, but choosing $g^{s}_{\chi}= 2 g^{p}_{\chi}$ ($g^{p}_{\chi}= 2 g^{s}_{\chi}$): 
for the same range as considered before for all other parameters, 
the relic density decreases in this case by at most $\sim$11\% ($\sim$2 \%) and the kinetic decoupling temperature increases (decreases) by up to $\sim$4\% ($\sim$12\%).

Allowing for a non-negligible coupling of the exchange particles to quarks, $g^s_q\neq0$, would introduce the possibility of further DM scattering processes $\chi q\leftrightarrow\chi q$  at high temperatures. However, since kinetic decoupling usually happens after the $SU(3)$ phase transition at $T_{\rm QCD}\sim170\,$MeV \cite{TQCD}, there are no free quarks around anymore and this does not affect our determination of $T_{\rm kd}$ -- unless one adopts rather large values for $m_\phi\gtrsim5\,$GeV
(if kinetic decouplings happens at $T\lesssim T_{\rm QCD}$,  the calculation of $T_{\rm kd}$ is also affected by the dependence of the effective number of relativistic degrees of freedom \emph{during} the transition; in our calculations, we used the values provided in Ref.~\cite{Hindmarsh:2005ix}).

We caution again that our results should be interpreted with care in some special regions of the parameter space where, due to the Ramsauer-Townsend effect, the assumption of a Maxwellian velocity distribution after kinetic decoupling may not be a good approximation anymore (see the discussion at the end of Sec.~\ref{dm_selfscatter}). 
The same goes for the calculation of $M_{\rm cut}$ in those strongly resonant cases where the asymptotic (``free'') behavior of the WIMP phase-space distribution is only reached after matter-radiation equality;
c.f.~the disclaimer at the end of Sec.~\ref{protohalos}. The fact that we find viable models where 
this indeed happens provides a strong motivation for future analyses to actually establish an exact relation between decoupling temperature and cutoff mass even during and slightly before matter domination -- corresponding to Eqs.~(\ref{mfs},\ref{mao}) which, strictly speaking,  only hold approximately in this case.

Let us finally briefly discuss possible ways to \emph{directly} probe the cutoff mass -- which would provide a fascinating new window into the particle nature of DM. While {gamma rays} (mostly via $\bar \chi\chi\rightarrow \ell^+\ell^-\gamma$) from individual subhalos  close to $M_{\rm cut}$ are unlikely to be resolved \cite{Pieri:2005pg}, the one-point probability function of the diffuse gamma-ray flux \cite{Lee:2008fm} could provide a better future probe; for very large cutoff masses, $M_{\rm cut}\gg M_\odot$, also other 
anisotropy probes could be sensitive enough  \cite{gammaanisotropy}.
 The last comment also holds for probing the smallest halos with gravitational lensing, especially for future astrometric microlensing missions with unprecedented sensitivities \cite{Li:2012qh}; 
in fact, it has been argued that even sub-solar objects could create observable strong gravitational lensing events, especially when making use of multiple images of time-varying sources \cite{Moustakas:2009na} (though this will be especially challenging in the case we are interested in here because the Einstein radius of DM subhalos is  much smaller than their virial radius).
For a review on  detectional prospects for sub-solar mass DM subhalos, see also Ref.~\cite{Koushiappas:2009du}.

Perhaps the most interesting \emph{indirect} probe of the cutoff scale, on the other hand, currently comes from the observation that galaxy clusters maximize the enhancement of a generic DM annihilation signal due to the presence of substructure  \cite{Pinzke:2009cp,Pinzke:2011ek}.
Assuming that one can extrapolate the results of numerical $N$-body simulations of gravitational clustering from the current resolution limit down to the much smaller cutoff values in the power spectrum that we are interested in here, it was shown in these references that small values of $M_{\rm cut}$ are strongly constrained by gamma-ray observations of galaxy clusters in models where the DM annihilation is Sommerfeld-enhanced; in the case of  leptophilic models, the main  source of gamma rays would be inverse Compton scattering of the high-energy leptons from $\bar\chi\chi\rightarrow\ell^+\ell^-$ off cosmic microwave or starlight photons. For a very specific model, with parameters chosen such as to result in a positron spectrum that could account for the cosmic ray excess ($m_\chi=1.6\,$TeV and $\phi\rightarrow\mu^\pm,e^\pm,\pi^\pm$ with a ratio of ~$\frac14:\frac14:\frac12$), the authors could show that for $M_{\rm cut}=10^{-6}M_\odot$ ($10^{4}M_\odot$) the maximally allowed Sommerfeld enhancement would be $S\sim5$ (800).\footnote{
This would lead to the conclusion that this or very similar leptophilic models cannot account for the cosmic ray lepton excess without violating the gamma-ray constraints from clusters if $M_{\rm cut}\leq10^4M_\odot$ \cite{Pinzke:2011ek}. While it is certainly not the prime purpose of our article to make such a connection, let us just point out that we could accommodate larger values of  $M_{\rm cut}$ even in our simple toy-model by choosing $m_s\ll100\,$MeV; a possibly desired decay into $\mu^\pm$ and $\pi^\pm$ could in that case exclusively happen through the (correspondingly heavier) pseudoscalar $\phi_p$.
}
 Here, $S=S(v\simeq124$km/s) and it was assumed that the annihilation cross section with Sommerfeld enhancement is given by $\sigma v=3\cdot10^{-26}$cm$^3$/s.

While a detailed analysis that extends the above results to more general models with Sommerfeld enhancements is certainly warranted, it is beyond the scope of this work. Here, we simply point out that we have presented a way to accurately calculate the cutoff mass in any given such model; limits on $M_{\rm cut}$ can thus directly be translated into limits on the model parameters (which determine the size of $S$). What is most interesting in this respect is that these limits will be strongest for small values of the lepton coupling $g_\ell^\phi$ because this leads to smaller values for $M_{\rm cut}$  -- which means that gamma-ray constraints from galaxy clusters probe the parameter space of leptophilic models from a completely different direction than other experiments (see Appendix \ref{app:Constraints} for a comparison).

Before passing to our conclusions let us finally mention yet another indirect probe of the small-scale cutoff that has  recently been suggested and may indeed turn out to be very promising: for both Kaluza-Klein  and neutralino DM, it was found in \cite{profumo_new} that the value of $M_{\rm cut}$ strongly correlates with the spin-dependent scattering rate of DM with nuclei, which is relevant both for direct detection experiments and for indirect DM searches looking for neutrinos from the sun. A corresponding signal would thus considerably narrow down the possible range of $M_{\rm cut}$ in these cases and it would be very interesting to see whether the same holds in models with large Sommerfeld enhancements.

\section{Summary and conclusions}
\label{sec:conc}

As already extensively discussed in the literature, the impact of Sommerfeld-enhanced annihilation rates on the DM relic density can be sizable both before \cite{Hisano:2006nn,Cirelli:2007xd,MarchRussell:2008yu,Iminniyaz:2010hy,Hryczuk:2010zi,Hryczuk:2011tq} 
and after \cite{Dent:2009bv,Zavala:2009mi,Feng:2010zp}
 kinetic decoupling. In this article, we have introduced a general framework that, for the first time, allows to \emph{consistently} describe situations where DM annihilation continues after chemical decoupling and interferes with kinetic decoupling, improving thus on the general praxis of using the kinetic decoupling temperature as an essentially free parameter. 
The coupled set of Eqs.~(\ref{Y_coupled}, \ref{y_coupled}) that describes the evolution of the WIMP number density and temperature thus provides one of the central results of this article.
As a consequence of our discussion, we have also refined the usual definition of kinetic decoupling in order to discriminate it more clearly from the point where WIMP (self-)interactions finally come to an end, which may  happen much later.

Applying our formalism to a simple leptophilic toy-model, we find that the impact of a new era of DM annihilation (i.e.~after kinetic decoupling) on the relic density can be significant. \emph{Off}-resonance, this effect is at most $\sim10$\% (at least in our model) and thus smaller than claimed in, e.g., Ref.~\cite{Dent:2009bv}. \emph{On}-resonance, on the other hand, we have demonstrated that DM annihilation can continue until well after matter-radiation-equality, depleting the DM abundance by more than two orders of magnitude after kinetic decoupling. This is a rather new result which completely changes the naive picture of associating the relic density of thermally produced DM to processes restricted to temperatures around
 $T_{\rm cd}\sim m_\chi/25$, i.e.~the very early universe.

Concerning the cutoff in the power spectrum of matter density fluctuations, we find that the resulting smallest DM protohalos form with masses in the range of roughly $M_{\rm cut}\sim \mathcal{O}(10^{-10}M_\odot)-\mathcal{O}(10\,M_\odot)$, depending on which experimental limits (see Appendix \ref{app:Constraints}) and which  model parameters  one chooses to adopt. As it turns out, the correctly determined cutoff mass can be almost two orders of magnitude larger than in the case where the impact of  DM annihilation after kinetic decoupling on the evolution of the DM phase-space distribution is not taken into account. In general, much smaller kinetic decoupling temperatures, and thus larger cutoff masses, are possible than for typical WIMPs like neutralino or Kaluza-Klein DM \cite{Profumo:2006bv,Bringmann:2009vf}, which may eventually even help to distinguish between these types of DM models (note that even $M_{\rm cut}\gg10\,M_\odot$ is possible for exchange particles lighter than  $100\,$MeV). Existing limits from gamma-ray observations of galaxy clusters \cite{Pinzke:2011ek} may already now be used to rule out the smallest values for the cutoff mass; since early kinetic decoupling happens for small lepton couplings,  this places limits on the parameter space that is complementary to bounds from $g-2$ measurements or beam dump experiments.

While we have chosen a specific, and rather simple, leptophilic toy-model for illustration, let us stress once again that our formalism can be used for \emph{any} model with annihilation rates that are enhanced at small velocities -- including more realistic models motivated by the cosmic ray anomalies (see e.g.~Ref.~\cite{CRrecent} for a recent discussion),
 more classical WIMPs like heavy neutralino DM \cite{Hisano:2006nn,Hryczuk:2010zi,Hryczuk:2011tq}
 or models where the annihilation rate is enhanced through the formation of bound states
 \cite{MarchRussell:2008tu}
 or an $s$-channel resonance
\cite{BWresonance}.
 In fact, in order to obtain reliable estimates for both the relic density and the small-scale cutoff of matter density perturbations in these cases, we have demonstrated here that it is \emph{mandatory} to use a framework that consistently takes into account the intertwined nature of WIMP annihilation into, and scattering with, heat-bath particles in the early universe.

\smallskip
\acknowledgments
We would like to thank Christoph Pfrommer for very useful communications concerning the gamma-ray limits from clusters and Christoph Weniger for insightful comments regarding the collision term.
L.v.d.A. and T.B. acknowledge support from the German Research Foundation (DFG) through the Emmy Noether Grant No.~BR 3954/1-1.

\appendix

\section{Sommerfeld enhancement}
\label{app:sommerfeld}

The Sommerfeld effect \cite{Sommerfeld:1931} arises when nonrelativistic DM particles interact via the exchange of force carriers $\phi$ that are much lighter than the DM particles themselves, $m_\phi\ll m_\chi$. Multiple $\phi$-exchanges  then result in nonperturbative corrections that can enhance both the DM annihilation  (see, e.g., \cite{Hisano:2002fk,Hisano:2003ec,Hisano:2004ds,Cirelli:2008jk,ArkaniHamed:2008qn,Pospelov:2008jd,Cholis:2008qq,Nomura:2008ru,Fox:2008kb,Lattanzi:2008qa}) and self-scattering \cite{Feng:2010zp, Buckley:2009in, Feng:2009hw} cross-section significantly.
An effective resummation over all corresponding ladder diagrams is performed by solving the Sch\"odinger equation for the two-body system,
\begin{equation}
	-\frac{1}{2 \mu} \nabla^{2} \psi_{k} = \left( \frac{k^{2}}{2 \mu} -V(r) \right) \psi_{k}\,,\label{eq:schreq}
\end{equation}
where $\psi_{k}$ is the full two-body wave-function, $k = m_{\chi} v$ is the momentum of each particle in the center-of-mass frame, $\mu = m_{\chi}/2$ is the reduced mass of the system and the potential $V(r)$ depends on the relative distance between the two particles $r$.

\subsection{Annihilation}

Because the potential typically has a much longer range than the very short relative distance $r$ at which the annihilation takes place, these two effects do not interfere: the only relevant effect of the potential is to change the free wave-function at $r=0$. The full annihilation rate is thus simply obtained by multiplying the unperturbed rate with the enhancement factor 
\be
	S(v) = \left| \psi_k(0) \right|^{2}\,, \label{eq:S}
\ee
where $\psi_k$ is the solution to \eref{eq:schreq} and  normalized as
\be	
	\psi \rightarrow \exp^{i k z} + f(\theta) \frac{\exp^{ikr}}{r} \quad \text{for } r \rightarrow \infty \,. 
\ee

For a massless force carrier the potential is Coulomb-like, $V(r) =- \alpha/r$, and the Schr\"odinger equation can be solved analytically to give
\begin{equation}
	S(v) = \frac{\pi /\epsilon_{v}}{1-e^{-\pi /\epsilon_{v}}}\,, \label{eq:SEana}
\end{equation} 
where $\epsilon_{v} \equiv v/\alpha$. For small velocities, $\epsilon_{v}\ll1$, the Sommerfeld enhancement thus is given by $S\simeq \pi /\epsilon_{v} $, while there will be no enhancement for large velocities, i.e. $S \approx 1 $ for $\epsilon_{v} \gg 1$. 

When the mediator has a nonzero mass $m_{\phi}$, the potential becomes Yukawa-like, $V(r) =- (\alpha/r)  e^{-m_{\phi}r}$, and \eref{eq:schreq} can only be solved numerically (see, e.g., Ref.~\cite{Hannestad:2010zt} for a pedagogic treatment). Since $V(r)$ is rotationally symmetric, we can expand $\psi_{k}$ into products of Legendre polynomials and radial functions $R_{kl}$. The only $R_{kl}$ that is nonzero at $r=0$ has angular momentum $l=0$, and therefore we substitute $R_{k0} \equiv \xi/r$. $\xi$ then obeys the radial Schr\"odinger equation
\be
\frac{1}{m_{\chi}} \frac{d^{2} \xi}{d r^{2}}=\left( - \frac{\alpha}{r} e^{- m_{\phi} r} - m_{\chi} v^{2} \right)  \xi \label{eq:radschreq}
\ee
with the following boundary conditions: 
\bea
\xi(0) &=& 0\,, \\
\frac{d \xi}{d r} &=&i m_{\chi} v \, \xi \quad \text{for } r \rightarrow \infty\, .
\eea

Expanding the Yukawa potential in powers of $m_{\phi}r$, $V\sim-\alpha/r + \alpha m_{\phi} +\mathcal{O}(r^{2})$, one recovers the Schr\"odinger equation with a Coulomb potential for \eref{eq:radschreq} when $\alpha m_{\phi}\ll m_{\chi} v^{2}$. Therefore one can safely use \eref{eq:SEana} as a good approximation to $S$ for a massive mediator in the range $\epsilon_{v}^2 \gg \epsilon_{\phi} \equiv {m_{\phi} / (\alpha m_{\chi}}) $.

For smaller velocities, the Coulomb approximation is no longer  valid, i.e. the exchange particle can no longer be treated as effectively massless. In this case, the finite range of the Yukawa potential induces resonances in $S$ that correspond to quasibound states of the pair of the incoming DM particles. We can get some more  insight in the behavior of these resonances by considering the Hulth\'en potential 
instead, which can be tuned to approximate the Yukawa potential very well and has the advantage to be analytically solvable \cite{Cassel:2009wt}. In this case, the resonances appear at specific values of $\epsilon_{\phi}\simeq6 (\pi n)^{-2}$, where $n$ is a positive integer, and the Sommerfeld enhancement at a resonance is given by $S \simeq (\pi^2/6)  \epsilon_{\phi}/\epsilon_{v}^{2}$ \cite{Slatyer:2009vg,Feng:2010zp}.
In reality, the Sommerfeld factor is limited by the finite lifetime of the bound state and therefore should not diverge as $v\rightarrow0$; rather, it is expected to saturate for \cite{Feng:2010zp,Hisano:2004ds} 
\be
\epsilon_{v} \lesssim \epsilon_v^{\rm cut, on} \equiv \alpha^{3} \epsilon_{\phi}\,. \label{eq:satonres}
\ee 
In our treatment, we therefore always use $S(\epsilon_{v}<\epsilon_v^{\rm cut, on})=S(\epsilon_v^{\rm cut, on})$.

Finally, we note that off the resonances,  the full numerical solution to $S$ saturates much earlier and stays constant for \cite{Lattanzi:2008qa}
\be
\epsilon_{v}  \lesssim \epsilon_v^{\rm cut, off} \equiv  0.5\,  \epsilon_{\phi}\,.\label{eq:satoffres}
\ee
This simply reflects the fact that the total effective energy seen by a Coulomb-like potential, $E_{\rm eff}\sim \alpha m_\phi + m_\chi v^2$ from the expansion of the Yukawa potential, essentially no longer depends on the velocity in this regime.


\subsection{Self-scattering}
\label{app:somm_scatt}

In the case of DM self-scattering, the repeated exchange of a gauge boson does not only affect the initial, but also the final-state wave-function; a separation between short- and long-distance scales is thus no longer possible and, in principle, all angular momentum contributions need to be taken into account. The radial Schr\"odinger equation describing self-scattering is thus the same as in the case of annihilation, Eq.~(\ref{eq:radschreq}), but with an additional centrifugal term $l(l+1)/(m_{\chi}r^{2})$ in the potential.
Its numerical solution  can be found by imposing appropriate boundary conditions, i.e. demanding that $\xi_{l}$ is regular at the origin and 
\be
	\xi_{l}\rightarrow \sin(m_{\chi}vr-\frac{\pi l}{2}+\delta_{l}) \qquad \text{for}\ r\rightarrow \infty\,.
\ee
From the phaseshift $\delta_{l}$, one can then calculate the partial scattering cross-section as
\be
	\sigma_{l} = \frac{4 \pi}{m^{2}_{\chi}v^{2}} (2 l+1) \sin^{2}{\delta_{l}}\,.
\ee 

For our purposes, we are mostly interested in the transfer cross section
\bea
\sigma_{T} &\equiv& \int d \Omega \ (1 - \cos \theta) \frac{d \sigma_{\chi\chi}}{d \Omega}\label{eq:transferxsection} \\
&=&\frac{4 \pi}{m^{2}_{\chi}v^{2}} \sum^{\infty}_{l=0}  \left[ (2 l+1) \sin^{2}{\delta_{l}} \right. \label{eq:transferxsection2}\\
	&&\left. \qquad -2(l+1)\sin{\delta_{l}} \sin{\delta_{l+1}} \cos{(\delta_{l+1}-\delta_{l})} \right] \,,\nonumber
\eea
which is a weighted integral over the differential scattering cross section. 
Estimating the maximum angular momentum $L$ that gives an important contribution to the sum, the authors of Ref.~\cite{Buckley:2009in} approximated the above full expression for $\sigma_T$ by assuming the phaseshift to be maximal for $l\leq L$, and minimal for $l>L$, i.e.~$\sin^{2}\delta_{l}=1$ and $0$ respectively. With this simplification, \eref{eq:transferxsection2} becomes 
\be
	\sigma_{\text{min}}=\frac{4 \pi}{m^{2}_{\chi} v^{2}} (1+L)\,, \label{eq:sigmamin}
\ee
which is in good agreement with the full numerical calculation \cite{Buckley:2009in}\footnote{
In \cite{Feng:2009hw, Feng:2010zp} the DM self-scattering rate was calculated in a different way, using known expressions for $\sigma_{T}$ to estimate the ion drag force in plasmas \cite{Khrapak:2003, Khrapak:2004}. The results of \cite{Feng:2009hw, Feng:2010zp} agree with \cite{Buckley:2009in} in the velocity regime that was of interest in these papers.}
 as long as $\epsilon_{\phi}\lesssim\epsilon_{v}\ll 1$; note that this condition is not satisfied in the Born regime (defined by large velocities, $\epsilon_v\gg1$, or small couplings, $\epsilon_\phi\gg1/2$).

For the models we consider in this work, we are particularly interested in very small velocities ($\epsilon_{v}\ll \epsilon_{\phi}$). In this case the cross section actually becomes velocity independent and the above approximation fails (unless we are near a resonance, see below): when $k R \ll1$ is satisfied ($R=m^{-1}_{\phi}$ being the effective range of the Yukawa potential), the phaseshift is given by
\be
	\delta_{l} \propto k^{2l+1} \,. \label{eq:delta_offres}
\ee  
All phaseshifts with $l \neq 0$ are thus negligible compared to $\delta_{0}$ and we have $\sigma_{T}\simeq \sigma_{l=0}$. The velocity independent \emph{effective potential cross section} is given by \cite{LandauLifshitz}
\be
	\sigma^{\rm pot} \equiv \sigma_{l=0}(kR\ll1) = 4 \pi a^{2} \,, \label{eq:sigma0}
\ee
where we used $\sin^{2}{\delta_{l}}\simeq \delta_{l}^{2}\equiv a^{2} k^{2}$, and $a$ is a constant called the scattering length.

As in the case of WIMP annihilation, resonances can occur for low-velocity scattering. Since the Schr\"odinger equation for $l=0$ scattering is the same as for annihilation, the resonances will occur at the same values of $\epsilon_{\phi}$. Assuming these quasi-bound states appear at an energy $E_{0}$ and have a width $\Gamma$, we are in the neighborhood of a resonance for $|E-E_{0}|\ll \Gamma$, with $E=m_{\chi} v^{2}$ the kinetic energy of the system. 
In this region, the approximation $\delta_{0}\simeq a k$ as given by \eref{eq:delta_offres} and \eref{eq:sigma0}  receives a considerable contribution from the resonance:
\be
	\delta^{\rm res}_{0} = \delta_{0} + \arctan{\left(\frac{\Gamma}{E_{0}-E}\right)}\,. \label{eq:delta_res}
\ee 
The full transfer cross section is thus given by
\be
	\sigma_{T}\simeq\sigma = \sigma^{\rm pot} +\sigma^{\rm res} \,, \label{eq:sigmatfull}
\ee
where the \emph{resonance scattering cross section} \cite{LandauLifshitz}
\be
	\sigma^{\text{res}} = \frac{4 \pi}{k^{2}} \frac{\Gamma^{2}-2 a k \Gamma (E-E_{0})}{(E-E_{0})^{2}+ \Gamma^{2}} \label{eq:sigmares}
\ee
is no longer negligible compared to $\sigma^{\rm pot}$, and should be taken into account. In fact, \eref{eq:sigmares} scales like $1/k^{2}$ and we can safely neglect $\sigma^{\rm pot}$ near a resonance. Exactly on-resonance, we  have
\be
	\sigma_{T} \simeq \sigma^{\rm res}(E=E_{0}) = \frac{4 \pi}{m^{2}_{\chi} v^{2}}\,,\label{eq:sigmatrres}
\ee
which is the same as \eref{eq:sigmamin} with $L=0$.

Far away from a resonance we cannot use \eref{eq:sigmamin} for $\sigma_{T}$ in the full velocity regime, since $\sigma^{\rm pot}$ becomes velocity independent as $v\rightarrow0$. In our analysis, we therefore use \eref{eq:sigmamin} down to the velocity at which $\sigma_T$ reaches the value of the asymptotic numerical solution for $\sigma_{l=0}(v\rightarrow0)$, beyond which we keep  $\sigma_T$ constant as in \eref{eq:sigma0}.
Note that around the transition point, at intermediate velocities, the full numerical solution actually results in $\sigma_T$ being significantly larger than $\sigma_{l=0}(v\rightarrow0)$ -- both because of the behavior of $\sigma_{l=0}$ and because of contributions from $l>0$; our approach is thus rather conservative in that it uses a  lower limit on $\sigma_{T}$.

For completeness, we note that  $\sigma$ can actually also exhibit ``anti-resonances'' for some values of $v$, and even (almost) completely vanish for very small velocities. This so-called\emph{ Ramsauer-Townsend effect }\cite{Ramsauer:1921} can be explained by a destructive interference between the potential and resonance scattering amplitude, cf.~\eref{eq:sigmatfull}; indeed, it follows from \eref{eq:sigmares} that the resonance contribution is negative for $\Gamma/(E-E_{0})<2 a k$. The cross section even disappears when the full phaseshift in \eref{eq:delta_res} becomes zero for $\Gamma/(E-E_{0})=\tan(a k)$. Since this only happens for special combinations of the parameters that appear in the Schr\"odinger equation, we do not take this effect into account in this work.

\section{Constraints on new light particles coupling to leptons}
\label{app:Constraints}
In this Appendix we summarize constraints on new light bosons $\phi$, in particular those of the type that appear in our toy-model introduced in Sec.~\ref{sec:toymod}. 
 
The strongest indirect constraints come from boson loops contributing to the anomalous magnetic moment of the muon, $a_\mu\equiv(g_\mu-2)/2$. The deviation between the experimentally measured and theoretically expected value is $\Delta a_\mu = a_\mu^{{\rm Exp}}-a_\mu^{{\rm SM}}= (255\pm63\pm49) \times 10^{-11}$, which corresponds to a discrepancy of $3.2$ times the estimated $1\sigma$ error  \cite{Nakamura:2010zzi}. We will require that the contribution of a new theory does not worsen this discrepancy beyond the $5\sigma$ level, i.e.~$-1.45\times10^{-9}\lesssim a_\mu^{\rm new}\lesssim 6.55\times10^{-9}$. 

The contribution of the pseudoscalar field to the magnetic moment of the muon is given by \cite{Leveille:1977rc}
\begin{equation}\label{mua}
a_\mu^p = \alpha_{\rm em}^{1/2}\,\frac{{g_\ell^p}^2}{4 \pi^{3/2}} \int_0^1 dx  \frac{-x^3}{x^2 + (1-x)m_p^2/m_\mu^2}
\end{equation}
and the contribution of the scalar field is given by 
\begin{equation}\label{mus}
a_\mu^s = \alpha_{\rm em}^{1/2}\,\frac{{g_\ell^s}^2}{4 \pi^{3/2}} \int_0^1 dx  \frac{2x^2-x^3}{x^2 + (1-x)m_s^2/m_\mu^2}.
\end{equation}
For a scalar of mass 0.1 (1, 10) GeV, the resulting limit on the coupling to muons is $g_\ell^s\lesssim1.8\times 10^{-3}$ ($6.7\times 10^{-3}$, $4.4\times 10^{-2}$); for pseudoscalars, the limit is stronger by a factor of roughly 2 (for similar constraints, albeit usually often considered for masses below 100 MeV, see e.g.~Refs.~\cite{Fayet:2007ua,Pospelov:2008zw, Freytsis:2009bh} and references therein).
Let us stress that these constraints only apply in the limit where the respective other particle's contribution  to $a_\mu$ can be neglected. Since the two contributions have opposite signs, however, in principle even  considerably larger values of  $g_\ell^s$ could be allowed for suitable choices of $m_p$ and $g_\ell^p$ if one is willing to accept the necessary fine-tuning. In the case of exactly degenerate  masses and couplings, e.g., the limits weaken to  $g_\ell^s=g_\ell^p\lesssim0.003$ ($0.016$, $0.15$) for  masses of $m_\phi=0.1$ (1, 10) GeV.

The most important direct constraints on new light bosons derive from beam dump experiments, where the incoming electrons could radiate such particles and one tries to spot their decay products behind the stopped electron beam
 (see, e.g., Ref.~\cite{Jaeckel:2010ni} for an overview and Refs.~\cite{Bjorken:2009mm,Freytsis:2009bh} for a recent discussion). The only such experiment that currently can probe scalar particles heavier than around 100 MeV, however, is E137 at Fermilab \cite{Bjorken:1988as}: for $100\,$MeV$\,\lesssim\!m_\phi\!\lesssim\!400\,$MeV, couplings to electrons are excluded in a rather narrow band of roughly $g_\ell^\phi\sim10^{-7}-10^{-6}$. 
Future beam dump experiments may extend such limits up to masses of a few GeV and close the gap to the limits obtained by  the muon $g-2$ constraints discussed above \cite{Bjorken:2009mm,Essig:2010xa,Abrahamyan:2011gv}.
New constraints in this region may possibly also be obtained with a low energy electron-proton collider \cite{Freytsis:2009bh}.

For masses $m_\phi>2\,m_\mu\approx210\,$MeV, the currently strongest constraints (apart from the small E137 window mentioned above) derive from the BaBar search for $\Upsilon$ decays into light (pseudo)scalar particles that subsequently decay into muons, $e^+e^-\rightarrow\Upsilon(3S)\rightarrow \gamma\phi$ and $\phi\rightarrow\mu^+\mu^-$ \cite{Aubert:2009cp}. Since the final states are identical, the BaBar limits can be interpreted as limits on the direct production of light bosons in $e^+e^-\rightarrow\gamma\phi$; assuming identical couplings to electrons and muons, this leads to roughly $g_\ell^\phi\lesssim10^{-3}$ -- with, however, considerably weaker constraints for masses around the $\rho$ resonance at $m_\phi\approx770\,$MeV \cite{Essig:2010xa}.

While our toy-model is only loosely motivated by the excess \cite{CR_excess} in cosmic ray leptons, let us mention that additional constraints in principle arise if one takes this connection seriously and requires that the positrons (and electrons) from DM annihilation do fit the cosmic ray data.
The fact that the cosmic ray antiproton spectrum \cite{Adriani:2010rc} is consistent with the expectation for the astrophysical background \cite{pbarbg} puts severe limits on hadronic decay modes of the new light particles \cite{Cirelli:2008pk,Donato:2008jk} compared to those required for the leptonic modes (see, e.g., Ref.~\cite{Bergstrom:2009fa}); as a consequence, the coupling to quarks should very roughly be suppressed as  $g^\phi_q\lesssim0.3\,g^\phi_\ell$. The large annihilation rates that are required to fit the cosmic ray lepton data are also potentially in conflict with both gamma-ray and radio observations towards the galactic center \cite{Bergstrom:2008ag,Baxter:2011rc}, gamma rays in the galactic halo due to inverse Compton scattering of the high-energy leptons  \cite{Meade:2009iu},
as well as with measurements of the cosmic microwave background \cite{Galli:2009zc,Slatyer:2009yq,Zavala:2009mi,Hannestad:2010zt} or even big bang nucleosynthesis \cite{Hisano:2011dc}
(see also Ref.~\cite{CRrecent} for an overview over these and similar constraints).

Completely independent of the cosmic ray connection, finally, very stringent constraints on \emph{low} values of $g_\ell$ arise in principle from DM annihilation-induced gamma rays from galaxy clusters since in this case the smallest protohalos form with very small masses and the annihilation flux from substructures is maximized; the corresponding constraints presented in Ref.~\cite{Pinzke:2011ek}, however, are rather model-dependent and not easily translated to our case -- see the dedicated discussion in Sec.~\ref{sec:disc}.

\section{Matrix elements and cross-sections}
\label{app:m2}

In this Appendix, we present relevant interaction rates for the leptophilic model introduced in Sec.~\ref{sec:toymod}.

The full expressions for the DM \emph{annihilation} cross section, which we used in our calculations, are somewhat lengthy and not very illuminating. Here, we therefore only state the result up to $\mathcal{O}\left(v^2,m_\phi^2/m_\chi^2\right)$:
\bea
  v_{{\rm rel}}\sigma_{\bar \chi \chi\rightarrow\phi_p\phi_s} &\simeq&  \frac{{g^s_{\chi}}^2 {g^p_{\chi}}^2}{16\pi m_\chi^2} 
  \Bigg[ 1+\frac14\frac {m_s^2+m_p^2} {m_\chi^2}  \\
  &&\phantom{\frac{{g^s_{\chi}}^2 {g^p_{\chi}}^2}{16\pi m_\chi^2}
  \Bigg[} -\left(3+\frac23 \frac{m_s^2+m_p^2} {m_\chi^2}\right)v^2 \Bigg]\,,  \nonumber \\
  v_{{\rm rel}}\sigma_{\bar\chi\chi\rightarrow\phi_s\phi_s} &\simeq& \frac{3{g^s_{\chi}}^4}{32\pi m_\chi^2}  \left[  1+\frac{11}{18}\frac {m_s^2}{m_\chi^2}\right]v^2\,,\\
v_{{\rm rel}}\sigma_{\bar\chi\chi\rightarrow\phi_p\phi_p} &\simeq& \frac{{g^p_{\chi}}^4}{96\pi m_\chi^2}  \left[  1-\frac{1}{2}\frac {m_p^2}{m_\chi^2}\right]v^2\,.
\eea
The fact that the two last cross-sections vanish for small velocities is simply a reflection of parity conservation. Note also that the direct $s$-channel annihilation into SM particles is strongly suppressed with $(g^{s/p}_\ell)^2$ and thus negligible.

The matrix element for the elastic \emph{scattering} of DM with standard model particles, if mediated only by  a scalar, is  given by
\begin{equation}
\label{chifchif}
{|\mathcal{M}^s|}^2= \frac{4{g^s_{\ell}}^2 {g^s_{\chi}}^2}{(t-m_s^2)^2}  \left( 4m_\chi^2-t \right) \left(4 m_{\ell}^2 -t \right)\,.
\end{equation}
In the case of a pseudoscalar mediator, we find
\begin{equation}
{|\mathcal{M}^p|}^2= \frac{4{g^p_{\ell}}^2 {g^p_{\chi}}^2}{(t-m_p^2)^2} ~t^2.
\end{equation}
In order to calculate the collision term for scattering processes to a sufficient accuracy, one only needs to evaluate the matrix elements at $t=0$, cf.~\eref{cTdef}. Therefore, only the scalar mediator gives a non-negligible contribution to the scattering processes we are interested in here.

The elastic scattering of DM with the light mediator particles $\phi$ can actually be much more efficient than the scattering with standard model particles -- at least as long as $n_\phi$ is not yet strongly Boltzmann suppressed. This can directly be seen from a comparison of \eref{chifchif} with the matrix element for $\chi\phi_s\leftrightarrow\chi\phi_s$ which, in the relevant limit, is given by
\be
  \mathop{\hspace{-12ex}{\left|\mathcal{M}\right|}^2_{t=0}}_{\hspace{4.5ex}s=m_\chi^2+2m_\chi\omega+m_s^2}=\frac{128\,{g_\chi^s}^4m_\chi^4\left(m_s^2-\omega^2\right)^2}{\left(m_s^4-4m_\chi^2\omega^2\right)^2}\,,
\ee
where $\omega$ is the energy of $\phi_s$. For the scattering with pseudoscalar particles, $\chi\phi_p\leftrightarrow\chi\phi_p$, this expression becomes
\be
  \mathop{\hspace{-12ex}{\left|\mathcal{M}\right|}^2_{t=0}}_{\hspace{4.5ex}s=m_\chi^2+2m_\chi\omega+m_p^2}=\frac{128\,{g_\chi^p}^4m_\chi^4\omega^4}{\left(m_p^4-4m_\chi^2\omega^2\right)^2}\,.
\ee
The inelastic scattering  $\chi\phi_p\leftrightarrow\chi\phi_s$ is generally suppressed with respect to the dominant of the above two scattering modes; for degenerate masses $m_\phi=m_p=m_s$, e.g., we find
\be
  \mathop{\hspace{-12ex}{\left|\mathcal{M}\right|}^2_{t=0}}_{\hspace{4.5ex}s=m_\chi^2+2m_\chi\omega+m_\phi^2}=\frac{32\,{g_\chi^p}^2{g_\chi^s}^2 m_\phi^4m_\chi^2\left(\omega^2-m_\phi^2\right)}{\left(m_\phi^4-4m_\chi^2\omega^2\right)^2}\,.
\ee

In order to apply our formalism even to scattering with the light mediator particles, we finally need  to make sure that these are still in thermal equilibrium with the heat bath. Unlike for the (assumedly) stable DM particles, equilibrium is in this case most efficiently maintained by (inverse) decay processes $\phi\leftrightarrow \bar\ell \ell$ and the first moment of the Boltzmann equation reads\footnote{
We also calculated the annihilation rates and found that these are, as expected, much smaller than the decay rates. In fact, the $\phi$ particles would in some cases \emph{never} have been in thermal equilibrium if only $2\leftrightarrow2$ processes were taken into account. 
}
\bea
  \label{boltz_phi}
 \dot n_\phi+3Hn_\phi &=& -\int\!\frac{d^3k}{(2\pi)^32\omega}\int\!\frac{d^3\tilde k}{(2\pi)^32\tilde \omega}\int\!\frac{d^3 p}{(2\pi)^3 2E}\nonumber\\
  &&\times(2\pi)^4\delta^{(4)}(p - \tilde k-k){\left|\mathcal{M}\right|}^2_{\phi\rightarrow  \bar\ell \ell}
  \nonumber\\
  &&
  \quad\Big\{\!\left[1- g^+(\omega)\right]\left[1- g^+(\tilde\omega)\right]\, f(\mathbf{p})
  \nonumber\\
  &&\qquad-g^+(\omega) g^+(\tilde\omega)\left[1+f(\mathbf{p})\right]\Big\} \label{boltzdecay1}\\
  &=& -\left<\Gamma\right>\left(n_\phi-n_\phi^{\rm eq}\right)\,.
\eea
Up to one remaining integral in $\omega_+\equiv\omega+\tilde{\omega}$, the above phase-space integrals can be fully performed analytically and the thermally averaged decay rate is given by
\bea
 \left<\Gamma\right>&=&  \frac{{\left|\mathcal{M}\right|}^2_{\phi\rightarrow  \bar\ell \ell}}{8\!\left(2\pi\right)^3\!n_\phi^{\rm eq}}  
\int\!\!\!\int \frac{d^3 k\, d^3\tilde k~ \delta(E-\omega-\tilde\omega)}{\omega\tilde\omega(\omega+\tilde\omega)\left(e^\frac{\omega}{T}+1\right)\left(e^\frac{\tilde\omega}{T}+1\right)}\nonumber\\
 \\
 &=&
 \frac{\Gamma_\phi T m_\phi}{\pi^2 n_\phi^{\rm eq} \sqrt{1-\frac{4m^2_\ell}{m_\phi^2}}} 
 \int_{m_\phi}^\infty\!\! 
\frac{d\omega_+}{e^\frac{\omega_+}{T}-1}\log\left[\frac{\cosh\frac{\omega_-^{\rm m}+\omega_+}{4T}}{\cosh\frac{\omega_-^{\rm m}-\omega_+}{4T}}\right]\,.\nonumber\\
\eea
In the above expressions, $E=\sqrt{(\mathbf{k}+\tilde{\mathbf{k}})^2+m_\phi^2}$, $\omega_-^{\rm m}=\sqrt{1-4m_\ell^2/m_\phi^2}\sqrt{\omega_+^2-m_\phi^2}$ and $n_\phi^{\rm eq}$ is the equilibrium number density which in the nonrelativistic limit becomes $n_\phi^{\rm eq}=m_\chi^2 T K_2(m_\phi/T)/(2\pi^2)$. For very small temperatures, $T\ll m_\phi$, $\langle\Gamma\rangle$ of course simply reduces to the decay rate at rest, for $\phi\rightarrow \bar\ell \ell$,   given by
\bea
  \Gamma_s &=& \frac{{g_\ell^s}^2}{8\pi}m_s\left(1-4\frac{m_\ell^2}{m_s^2}\right)^\frac32\,,\\
  \Gamma_p &=&  \frac{{g_\ell^p}^2}{8\pi}m_p\left(1-4\frac{m_\ell^2}{m_p^2}\right)^\frac12\,.
\eea

\newpage

\end{document}